\documentclass[%
 reprint,
 amsmath,amssymb,
 aps,
]{revtex4-2}

\usepackage{graphicx}
\usepackage{dcolumn}
\usepackage{bm}
\usepackage[version=4]{mhchem}
\usepackage{xcolor}

\begin{document}

\preprint{17072025-Bhattacharjee}

\title{Elucidating the impact of point defects on the structural, electronic, and mechanical behaviour of chromium nitride}

\author{Barsha Bhattacharjee}
 \affiliation{Advanced Research Center for Nanolithography, Science Park 106, 1098 XG Amsterdam, The Netherlands}
 \altaffiliation{Institute for Theoretical Physics, University of Amsterdam, Postbus 94485, 1090 GL Amsterdam, the Netherlands}
\author{Emilia Olsson}
 \email{k.i.e.olsson@uva.nl}
 \affiliation{Advanced Research Center for Nanolithography, Science Park 106, 1098 XG Amsterdam, The Netherlands}
 \altaffiliation{Institute for Theoretical Physics, University of Amsterdam, Postbus 94485, 1090 GL Amsterdam, the Netherlands}

\date{17 July 2025}

\begin{abstract}
Defect engineering offers an important route to property tuning of nanostructured coatings for advanced applications. Transition metal nitrides, such as CrN, are widely used for their mechanical resilience, but their nitrogen-rich analogue \ce{CrN2} remains poorly understood, especially at the atomic scale. This study employs density functional theory to investigate the energetics as well as how intrinsic defects (vacancies, interstitials, and anti-sites) and extrinsic impurities (hydrogen and oxygen) influence the structural, electronic, magnetic, and mechanical response of \ce{CrN2}, in comparison to the more commonly studied CrN. With directional N-N bonding and semiconducting character, \ce{CrN2} shows high sensitivity to defect incorporation, including local spin polarisation, gap states, and mechanical softening. In contrast, CrN’s metallic character enables effective screening of similar defects, preserving its structural, magnetic, electronic and mechanical integrity. However, hydrogen induces anisotropic distortions and mechanical degradation in CrN, while oxygen enhances hardness. These findings reveal how defect chemistry and bonding anisotropy govern mechanical performance, with implications for nanoscale control in coatings design.
 
\end{abstract}

\maketitle


\section{Introduction}
Chromium nitrides (Cr\textsubscript{x}N\textsubscript{y}) are a class of binary transition-metal nitrides (TMNs) well-known for hardness\cite{Hones1997CharacterizationCoatings,Wuhrer2004ACoatings,Fuentes2005RecentProtection}, resistance to wear and corrosion\cite{Holleck1995MultilayerProtection,Senf1999WearCoatings,Jagielski2000EffectSteel,Bertrand2000ACoatings,Schauperl2008WearCoatings}, high melting point\cite{Evseev2020ChromiumMode}, thermal stability\cite{Hultman2000ThermalFilms}, and electrical stability\cite{Eklund2016Transition-metal-nitride-basedMaterials},  with superior surface oxidation properties as compared to the other TMNs\cite{Esaka1997ComparisonSpectroscopy, Chim2009OxidationArc}. These properties make them suitable as hard protective coatings \cite{Vera2011ASubstrates,Qi2013ACoatings,Du2020MechanicalCoatings}, biomedical implants\cite{Rahman2020CorrosionApplications}, cutting tools\cite{Santhanam1996ApplicationTools}, in steel alloys\cite{Shabashov2022MechanosynthesisNitrides} and engine components\cite{Singh2020TribologicalTemperature}. Despite their intrinsic material advantages, the overall performance of many TMNs is influenced by the presence and nature of atomic-scale defects. Synthesis of stable sub-stoichiometric TM mononitrides has been reported for TiN \cite{Jiang1993Non-stoichiometryDeposition,Shin2003VacancyLayers,Shin2004GrowthMgO001,Vissers2013CharacterizationGrowth}, VN \cite{Skala1990NitrogenNitride, Gueddaoui2006EffectsFilms}, and CrN\cite{Gall2002GrowthProperties,Hones2003StructuralFilms,Subramanian2012InfluenceSputtering,Zhang2013InsightsCrN}. The recent discovery of nitrogen rich transition metal pernitrides (TMN\textsubscript{2}), which exhibit exceptional mechanical properties such as ultra-incompressibility and high hardness, has expanded the class of materials considered for use in industrial hard coatings \cite{Gregoryanz2004SynthesisNitride, Niwa2014Discoverysub2/sub, Liu2014ElectronicPernitrides, Bhadram2016High-PressurePernitride, Chen2017TheoreticalPressure, Zhang2019UnexpectedHf, Niwa2019High-pressureDimer}. While research has focused on these materials' stability and mechanical properties, understanding of the presence and effects of defects within TM pernitrides remains unexplored. Moreover, the performance of these materials as protective coatings is further influenced by the material's interaction with impurities like hydrogen and oxygen that can be present during their growth and become trapped at intrinsic defects sites, or absorbed from the atmosphere.\cite{Tsetseris2007StructureNitrides, Tsetseris2008TrappingStudy} These atomic impurities are particularly critical because they can lead to degradation mechanisms such as embrittlement\cite{Lee1996HydrogenCoatings}, compromising the material's performance and longevity. Therefore, it is crucial to understand how intrinsic and extrinsic point defects (impurities) influence the physical properties of CrN-based materials. \par

Several experimental and theoretical studies have reported changes in structural, electronic and mechanical properties in the presence of defects in TMNs. Experimental studies supported by density functional theory (DFT) calculations showed structural changes in MoN\cite{Ozsdolay2017CationNitride} where the lattice parameter was reduced with a decrease in cation to anion ratio, and in HfN\cite{Hu2015NegativeFilms} where the lattice distorted from cubic to rhombohedral. Similarly, changes in electronic properties such as the introduction of new states around the Fermi level was observed in ScN through experimental\cite{Cetnar2018ElectronicSubstrates} and theoretical \cite{Kumagai2018PointPrinciples, Deng2015OpticalGap, Kerdsongpanya2012EffectProperties} works, in TiN  experimentally \cite{Zhang2021CorrelatingFilms} and theoretically \cite{Tsetseris2007StructureNitrides,Lee2012Environment-dependentVacancies,Jhi2001VacancyNitrides, Shin2003VacancyLayers} and in VN experimentally \cite{Gueddaoui2006EffectsFilms, Aissani2021InvestigatingSputtering}. In terms of mechanical properties, an increase in the concentration of nitrogen vacancies was experimentally observed to reduce the elastic moduli of group 4 TMNs such as TiN\textsubscript{x} and ZrN\textsubscript{x} \cite{Holleck1986MaterialCoatings, Jiang1991ElasticIndentation, Shin2003VacancyLayers,KarimiAghda2023ValenceNitrides}, while in HfN\textsubscript{x}, the moduli first increase for $0.8 \leq x \leq 1.0$, and then decrease for $1.0 \leq x \leq 1.2$\cite{Hu2015NegativeFilms}. The opposite trend was observed for group 5 TMNs, where both VN\textsubscript{x} and NbN\textsubscript{x} have shown an increase in the elastic moduli\cite{Zhang2016GrowthMgO001, KarimiAghda2023ValenceNitrides}. Of the group 6 TMNs, studies show that vacancies stabilize the mechanically unstable MoN\textsubscript{x}\cite{Ozsdolay2016CubicRatio, Balasubramanian2017Phase1} and WN\textsubscript{x} \cite{Balasubramanian2016Vacancy-inducedNitride}. Yet, similar investigations of defect-property relationship into chromium-based nitrides remain limited, making the role of stoichiometry in determining their physical properties an open question. \par
Due to its structural stability and magnetic ordering, the chromium mononitride phase (CrN) is a key candidate for both fundamental studies and practical applications among \ce{Cr_xN_y}. CrN undergoes a magneto-structural transition, adopting a paramagnetic cubic (\textit{Fm$\overline{3}m$}) structure at room temperature and transitioning to an antiferromagnetic orthorhombic (\textit{Pnma}) phase near the Néel temperature (273-283 K) \cite{Browne1970AnCrN, Ibberson1992TheCrMoN, Gall2002GrowthProperties, SubramanyaHerle1997SynthesisCr2S3, Corliss1960AntiferromagneticCrN}. The experimentally measured mechanical hardness of CrN varies significantly ($\approx$ 11-26 GPa) depending on several factors including method of synthesis, microstructure, and residual stress \cite{Hones1997CharacterizationCoatings,Engel1998CorrosionDeposition, Cunha1999MicrostructureTechniques,Bertrand2000ACoatings}. Similarly, theoretical predictions span a broad range ($\approx$ 11-26 GPa), largely influenced by the choice of computational method and hardness model\cite{Tian2012MicroscopicCrystals, Kvashnin2017ComputationalMaterials}. In the presence of intrinsic defects, a combined experimental and theoretical study revealed that increasing nitrogen vacancies reduces the lattice constant, shifts the bonding character, and enhances metallic behaviour in cubic CrN  \cite{Zhang2013InsightsCrN}. A theoretical study \cite{Mozafari2015RolePrinciples} showed that nitrogen point defects, in the form of vacancies and interstitials, can close the non-zero band gap in semiconducting cubic CrN. This contrasts with experimental reports of pristine CrN exhibiting metallic behaviour across the transition and showing decreased resistivity below the Néel temperature \cite{Browne1970AnCrN, Tsuchiya1996Non-stoichiometrySputtering, Inumaru2007ControllingGrowth}. In the orthorhombic phase, intrinsic defects, including chromium and nitrogen vacancies, antisites, and interstitials, alter the electronic structure by introducing shallow states near the Fermi level, modifying \textit{d} and \textit{p} orbital distributions, inducing spin polarization, and disrupting antiferromagnetic order \cite{Rojas2018EnergeticsCrN}. Extrinsic defects such as oxygen incorporation increase CrN’s hardness \cite{Suzuki2006InfluenceDeposition,Du2021InfluenceCoating} but the effect of hydrogen remains unexplored. \par
The pernitride phase - CrN\textsubscript{2} crystallizes in the \textit{P6\textsubscript{3}/mmc} space group and features an anti-NiAs type structure. It consists of CrN\textsubscript{6} triangular prisms stacked along the c-axis through rigid N-N dimers, which contribute to its axial incompressibility making it harder than the mononitride phase \cite{Niwa2019High-pressureDimer}.  Early calculations suggested a Vickers hardness of approximately 46 GPa, indicating potential superhard characteristics \cite{Zhao2016PotentiallyPressure}. However, subsequent analyses reported an ideal shear strength around 30 GPa \cite{Liu2019TheStudy}, which is below the expected threshold for superhard materials. While Vickers hardness reflects resistance to plastic deformation from indentation, ideal shear strength provides a theoretical upper limit for shear deformation in defect-free crystals. The difference between the two highlights the role of atomic bonding and deformation mechanisms in governing hardness. These discrepancies underscore the need for comprehensive evaluations when assessing material hardness. Moreover, experimental approaches often face challenges in isolating and characterizing specific defects at the atomic level, making it difficult to establish a direct correlation between individual defects and their effect on material properties.  \par
To further understand the role of intrinsic defects and atomic impurities in the mechanical robustness of chromium nitride, we explore the influence of defects on compositions used in protective hard coatings using DFT. Specifically, we evaluate the mechanical properties of mononitride CrN $(x=y=1)$ and pernitride CrN\textsubscript{2} $(x=1, y=2)$, identifying them as hard materials with theoretically predicted Vicker’s hardness in the range typical for protective applications (below 40 GPa) \cite{Tian2012MicroscopicCrystals, Kanyanta2016HardOverview}. Focusing on both intrinsic defects (vacancies, anti-sites, interstitials) and extrinsic impurities (hydrogen and oxygen), we calculate the formation energies of these defects and investigate their effects on electronic and mechanical properties. This approach enables us to assess how such imperfections impact elastic moduli and hardness, thereby affecting the durability and performance of the mononitride and pernitride phases. By detailing the defect-related influences on structural and mechanical properties, this study aims to guide improvements in synthesis and treatment protocols that mitigate degradation, providing a pathway toward optimizing chromium nitride-based coatings.

\section{Methodology}
All calculations were conducted within the DFT framework using the Vienna Ab-Initio Simulation Package (VASP version 6.5.1) \cite{Kresse1993iAbMetals,Kresse1996EfficiencySet}. The Perdew-Burke-Ernzerhof (PBE) functional \cite{Perdew1996GeneralizedSimple} with the generalized gradient approximation (GGA) \cite{Langreth1983BeyondProperties} and the projector-augmented wave (PAW) method \cite{Blochl1994ProjectorMethod} were used to accurately model electronic exchange-correlation interactions and electron-ion interactions in our DFT calculations. To describe the strong on-site Coulomb interactions due to Cr \textit{3d} electrons, we applied the Dudarev approach with a Hubbard U term, \( U\textsubscript{eff} \) \cite{Anisimov1991BandiI/i} to CrN. DFT + U was not applied to CrN\textsubscript{2}, as test calculations showed that it led to significant overestimation of the lattice parameters compared to experimental values. After convergence tests on conventional unit cells (with convergence criteria of 1 meV per atom) the plane wave energy cutoff was set to 770 eV  and a $\Gamma$-centered k-point mesh of 16x16x16 for CrN and 14x14x5 for CrN\textsubscript{2} , with Methfessel-Paxton smearing\cite{Methfessel1989High-precisionMetals} with a width of 0.1 eV. Electronic convergence was achieved with energy changes below \(10^{-6} \, \text{eV}\), and ionic relaxation continued until forces on atoms were below \(-0.01 \, \text{eV/\AA}\). Different magnetic configurations, including antiferromagnetic (AFM) and ferromagnetic (FM) orderings, were tested for cubic CrN. The FM configuration was chosen as the representative magnetic state for further analysis (see Supplementary Information, Section 1). 

To assess the mechanical properties, elastic constants were calculated using the energy-strain method as implemented in VASPKIT \cite{Wang2021VASPKIT:Code}. Mechanical stability of these structures was assessed according to the Born-Huang stability criteria \cite{Born1996DynamicalLattices, Mouhat2014NecessarySystems}. Elastic properties were derived using the Voigt approximation, from which bulk modulus (K\textsubscript{V}), Young’s modulus (E\textsubscript{V}), and shear modulus (G\textsubscript{V}) were obtained. Polycrystalline averages were then used to evaluate bulk-relevant mechanical metrics, including Poisson’s ratio ($\nu$), Pugh’s ratio (k), the Universal Elastic Anisotropy index (A\textsuperscript{U}), and Vickers hardness (H\textsubscript{V}) estimated using Tian’s empirical model \cite{Tian2012MicroscopicCrystals}.

Defects were introduced into relaxed, pristine structures and subsequently geometry optimised. Supercell convergence tests were performed separately for CrN and CrN\textsubscript{2}. For CrN, cells ranging from 2×2×2 (64 atoms) to 4×4×4 (512 atoms) were tested, while for CrN\textsubscript{2}, supercell sizes from 2×2×2 (48 atoms) to 4×4×2 (192 atoms) were examined, including intermediate sizes such as 3×3×1 (54 atoms), 3×3×2 (108 atoms), and 3×3×3 (162 atoms). Convergence was defined by a change in defect formation energy of less than 0.02 eV between successive supercells. To minimize size effects while maintaining computational efficiency, a 3x3x3 supercell (216 atoms) was selected for CrN, and a 3x3x2 supercell (108 atoms) for CrN\textsubscript{2}. The corresponding k-point meshes were 2x2x2 for CrN and 5x5x3 for CrN\textsubscript{2}. Inequivalent vacancy and antisite defect sites were identified using the Site-Occupation Disorder (SOD) code \cite{Grau-Crespo2007Symmetry-adaptedSolids}, while interstitials were located via Atomsk \cite{Hirel2015Atomsk:Files} or identified by symmetry-based crystallographic sites (tetrahedral, octahedral) from the Bilbao Crystallographic Server \cite{Aroyo2006BilbaoGroups}. The defect formation energy, \( E_f^{def}(q) \), is defined as \cite{Zhang1991ChemicalSelf-diffusion}:
\begin{equation}
\label{Edef}
    E_{f}^{def}(q) = E_{def} - E_{f}^{bulk} - \sum_{i} n_{i}\mu_{i} + q(E_{f} + E_{VBM}) + E_{corr}
\end{equation}
where \( E_{def} \) is the total energy of the defective cell, \( E_{bulk} \) is the total energy of the pristine cell, and \( \mu_i \) is the chemical potential of species \( i \) added or removed. \( E_f \) represents the Fermi energy relative to the valence band maximum, \( E_{VBM} \). All defect energetics reported herein follow Kröger-Vink notation \cite{Kroger1956RelationsSolids}. The stability of point defects was evaluated under both Cr-rich and N-rich conditions \cite{Buckeridge2014AutomatedCompounds}. These conditions were defined through chemical potentials of Cr (\( \mu_{Cr} \)) and N (\( \mu_{N} \)), which vary depending on the growth environment. Under Cr-rich conditions \( \mu_{Cr} \) is set equal to the chemical potential of bulk Cr, while \( \mu_{N} \) is derived from the equilibrium condition \( \mu_{Cr} +\mu_{N}=\mu_{CrN}\), where \( \mu_{CrN} \) is the chemical potential of bulk CrN. Conversely, under N-rich conditions, \( \mu_{N} \) is set equal to half the chemical potential of molecular nitrogen gas (\( \mu_{N} =\frac{1}{2}\mu_{N_2}\)), and \( \mu_{Cr} \) is again determined by equilibrium with bulk CrN.  For materials with a band gap, a correction term \( E_{corr} \) was applied to address electrostatic interactions between charged defects and their periodic images. The Kumagai-Oba scheme \cite{Kumagai2014Electrostatics-basedCalculations}, as implemented in Spinney \cite{Arrigoni2021Spinney:Python} was employed, as it refines the Freysoldt, Neugebauer, and Van de Walle (FNV) approach \cite{Freysoldt2009FullyCalculations,Freysoldt2011ElectrostaticSupercells,Freysoldt2014First-principlesSolids} by using anisotropic point-charge energy and atomic-site potentials for alignment. This is particularly important for systems like \ce{CrN2}, which exhibit anisotropic dielectric tensors due to their non-cubic symmetry.  For CrN\textsubscript{2}, which has a calculated band gap of 0.73 eV, the dielectric constant was calculated via density functional perturbation theory (DFPT) as $\epsilon_{xx} = 26.085$ and $\epsilon_{zz} = 22.378$, allowing for defect-related changes in physical properties to be assessed for the most stable configurations. All analysis in the main text focuses on neutral defects to provide a consistent, Fermi-level-independent reference for comparing local bonding and electronic structure across systems. Charged defect configurations, which depend on Fermi level and dielectric screening, are then discussed in the Supplementary Information (Figure S8).

\section{Results}
The structural and physical properties of \ce{Cr_xN_y} were initially examined across five stoichiometric compositions to assess trends in stability and defect behaviour. Among these, CrN ($x=1, y=1$) is known to be the most thermodynamically stable phase, consistent with previous reports \cite{Yan2014StructuralStudy,Kvashnin2017ComputationalMaterials}, and it serves as a reference point in this study. CrN\textsubscript{2}, by contrast, appears as a mechanically hard metastable phase within the present simulations. Further details on the stability of the individual materials and their physical properties can be found in the Supplementary Information (Section 1). Defects lead to lattice distortions, changes in the electronic landscape and mechanical properties. We describe the effects of point intrinsic defects and impurities in the form of oxygen and hydrogen atoms on the physical properties of the two mechanically hard phases of Cr\textsubscript{x}N\textsubscript{y}: CrN\textsubscript{2} and CrN. The defects are grouped into the following sets: (i) chromium defects: vacancy, antisite and interstitial (ii) nitrogen defects: vacancy, antisite and interstitial (iii) oxygen defects: substitution at both chromium and nitrogen site and (iv) hydrogen defects: substitution at both chromium and nitrogen site. We then identify the most probable defects based on their defect formation energies, and discuss their impact on the physical properties as compared to their pristine bulk counterparts. 
\par
\subsection{Defect Energetics, Structure, and Magnetic Response}

The lowest defect formation energies ($E_{f}^{\text{def}}(q)$) for various defects, calculated using equation~\ref{Edef}, are presented in Table~\ref{table1}. These correspond to the most stable configurations, identified by sampling multiple atomic sites and charge states across the Fermi level. The results are shown for both N-rich (Cr-poor) and N-poor (Cr-rich) chemical potential limits, with the Fermi level fixed at the valence band maximum (VBM). Results of alternative configurations and charge states are provided in the Supplementary Information. Under both N-rich (Cr-poor) and N-poor (Cr-rich) conditions, $V_\text{N}$ is the most favourable intrinsic defect in both CrN and CrN\textsubscript{2}, with neutral formation energies of 1.71 eV and 2.34 eV, respectively. The enhanced stability in CrN is attributed to the absence of N-N dimer bonding found in CrN\textsubscript{2}, which makes $V_\text{N}$ more disruptive in the pernitride. Nitrogen anti-sites and N interstitials in CrN\textsubscript{2} are highly unfavourable ($E_f^\text{def}(q) >$ 5 eV), consistent with the tight bonding environment in the dimerized structure. CrN, in contrast, stabilizes a split N interstitial at 3.41 eV, close to prior reports using different methodologies \cite{Mozafari2015RolePrinciples}. Most Cr-related defects in both systems exhibit high formation energies ($>$5 eV), with the exception of the Cr vacancy in CrN ($V_\text{Cr} = 2.13$ eV), suggesting moderate feasibility. \par

\begin{table}[h]
\small
\caption{$E_f^\text{def}$ values of the most stable defects in Cr\textsubscript{x}N\textsubscript{y}.}
\label{table1}
\begin{tabular*}{0.48\textwidth}{@{\extracolsep{\fill}}lllll}
\hline
\textbf{Composition} & \textbf{Defect} & \textbf{N-poor} (eV) & \textbf{N-rich}(eV) \\
\hline
\multicolumn{4}{l}{\textbf{CrN\textsubscript{2}}} \\
&$V_{\text{Cr}}^\times$     & 6.10 & 5.06 \\
&$V_{\text{N}}^\times$     & 1.82 & 2.34 \\
&$\text{H}_{\text{N}}^\times$ & 0.30 & 1.50 \\
&$\text{O}_{\text{N}}^\times$ & -0.46 & 0.06 \\
\multicolumn{4}{l}{\textbf{CrN}} \\
&$V_{\text{Cr}}^\times$     & 3.30 & 2.13 \\
&$V_{\text{N}}^\times$     & 0.54 & 1.71 \\
&$\text{H}_{i}^\times$     & 1.17 & 1.17 \\
&$\text{O}_{\text{N}}^\times$ & -3.39 & -2.81 \\
\hline
\end{tabular*}
\end{table}

Extrinsic defects show stronger chemical potential dependence. In both CrN\textsubscript{2} and CrN, substitutions at the chromium site, $H_{Cr}$ and $O_{Cr}$, are energetically unfavourable ($E_f^\text{def}(q) >$ 5 eV), whereas N-site substitutions and some interstitials are stable. In CrN\textsubscript{2}, $\text{H}_\text{N}^\times$ has a formation energy of 1.50 eV and exhibits charge transitions at 0.38 eV and 0.52 eV. Two low-energy interstitial H configurations at C\textsubscript{3v} symmetry yield values of 2.58 eV and 2.92 eV. In CrN, the most stable H defect is the interstitial site($\text{H}_i^\times = 1.17$ eV), followed by the bond-centred site (1.38 eV) and the tetrahedral site (2.32 eV). Hydrogen substitution at the N site ($\text{H}_\text{N}^\times = 1.36$ eV) is also thermodynamically accessible. Oxygen incorporation follows similar trends. In CrN\textsubscript{2}, $\text{O}_\text{N}^\times = 0.06$ eV with a charge transition at 0.63 eV. In CrN, $\text{O}_\text{N}^\times = -2.81$ eV, indicating a strong thermodynamic driving force due to the greater stability of Cr-O bonds compared to Cr-N \cite{Wang2022CompositionFilms}. Oxygen interstitials are generally unfavourable in CrN\textsubscript{2} ($E_f^\text{def}(q) >$ 5 eV), while CrN supports three configurations below 5 eV, including a split interstitial ($E_f^\text{def}(q) = 1.35$ eV, C\textsubscript{s} symmetry), and two higher-energy sites (3.47 and 3.91 eV), illustrated in the Supplementary Information (Figure S5).

\begin{figure*}
        \includegraphics[width=0.8\textwidth]{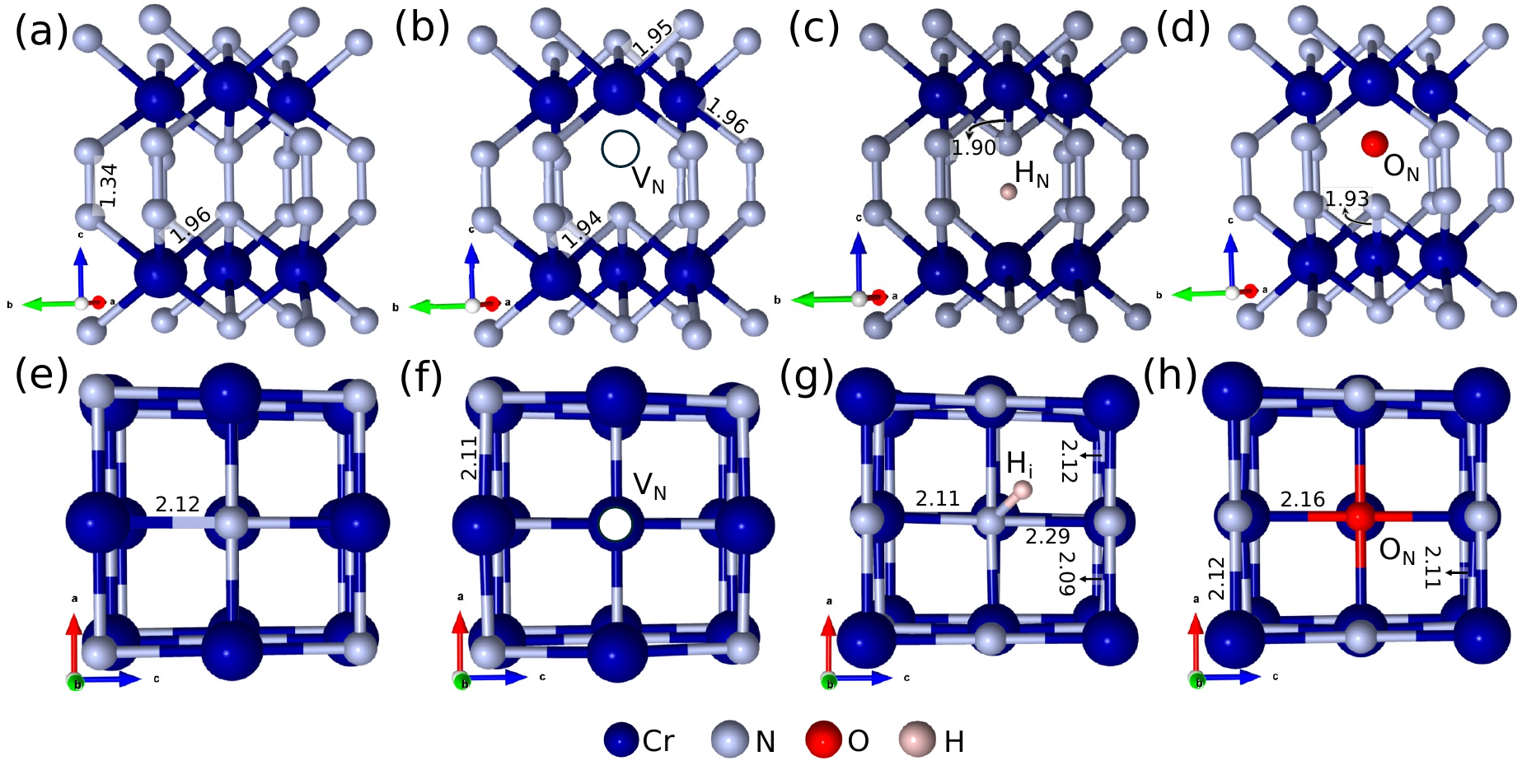}
        \caption{\label{fig1} Overview of structural distortions in presence of defects. 1(a) - (d) show the pristine, $V_{\text{N}}^\times$, $\text{H}_{\text{N}}^\times$, $\text{O}_{\text{N}}^\times$, respectively in CrN\textsubscript{2} and (e) - (f) show the pristine, $V_{\text{N}}^\times$, $\text{H}_{i}^\times$ and $\text{O}_{\text{N}}^\times$ in CrN.
        }
\end{figure*}

Point defects perturb the local bonding environment, introducing structural distortions that often couple with changes in charge and spin distribution. Figure\ref{fig1} presents relaxed structures of the lowest-energy neutral point defects, with bond length distributions provided in the Supplementary Information (Figure S6, S7). In CrN\textsubscript{2}, a nitrogen vacancy ($V_\text{N}$) leads to slight bond contraction (0.01-0.02 \text{\AA}) in nearby Cr-N and N-N pairs, attributed to the removal of electronic repulsion from the missing N atom. In contrast, substitutional hydrogen ($\text{H}_\text{N}$) and oxygen ($\text{O}_\text{N}$) defects cause surrounding Cr and N atoms to move outward. In CrN, the $V_\text{N}$ defect displaces nearby Cr atoms away and N atoms toward the vacancy, reflecting the material’s metallic bonding and more delocalized electrons. Oxygen substitution ($\text{O}_\text{N}$) induces an asymmetric relaxation where Cr moves away and N moves closer to the defect site. Notably, $\text{O}_\text{N}$ aligns with the original N lattice site, suggesting minimal off-site relaxation while $\text{H}_i$ settle into off-centre positions, displacing adjacent Cr and N atoms due to their small size and high mobility. \par

CrN\textsubscript{2} is calculated to exhibit a non-magnetic ground state in its pristine form. Upon introduction of a $V_\text{N}$ defect, a total magnetic moment of 0.375~$\mu_B$ emerges, predominantly on adjacent Cr atoms, due to defect-induced electronic localisation. The origin and nature of these spin-polarised states are analysed in detail in the subsequent section. The $\text{H}_\text{N}$ defect does not alter the magnetic character, whereas $\text{O}_\text{N}$ introduces a magnetic moment of -1.041 $\mu_B$, again centred on nearby Cr atoms. This indicates a strong coupling between defect type, local structure, and spin polarization. In CrN, each Cr atom exhibits a magnetic moment of 2.97 $\mu_B$, and this local moment persists across all defect configurations examined. Although no significant net magnetization change is observed, the spin density redistributes locally, indicating defect-induced polarization without quenching the overall magnetic ordering.\par

These results reveal that the energetics and structural signatures of point defects in CrN\textsubscript{2} and CrN are highly sensitive to both defect chemistry and bonding environment. The formation of vacancies and substitutional impurities induces local atomic rearrangements that are more pronounced in CrN\textsubscript{2}, consistent with its directional N-N bonding. In contrast, CrN exhibits greater structural rigidity and greater tolerance to point defects, reflecting its more delocalised metallic bonding. Defect-induced spin polarisation is similarly material-dependent: while CrN\textsubscript{2} shows local magnetic moments in response to certain defects, CrN maintains its intrinsic magnetic character. These differences suggest fundamentally distinct mechanisms of defect accommodation in the two compounds, with potential implications for their electronic structure, as explored in the following section.

\subsection{Electronic properties}
\begin{figure*}
       \includegraphics[width=0.8\textwidth]{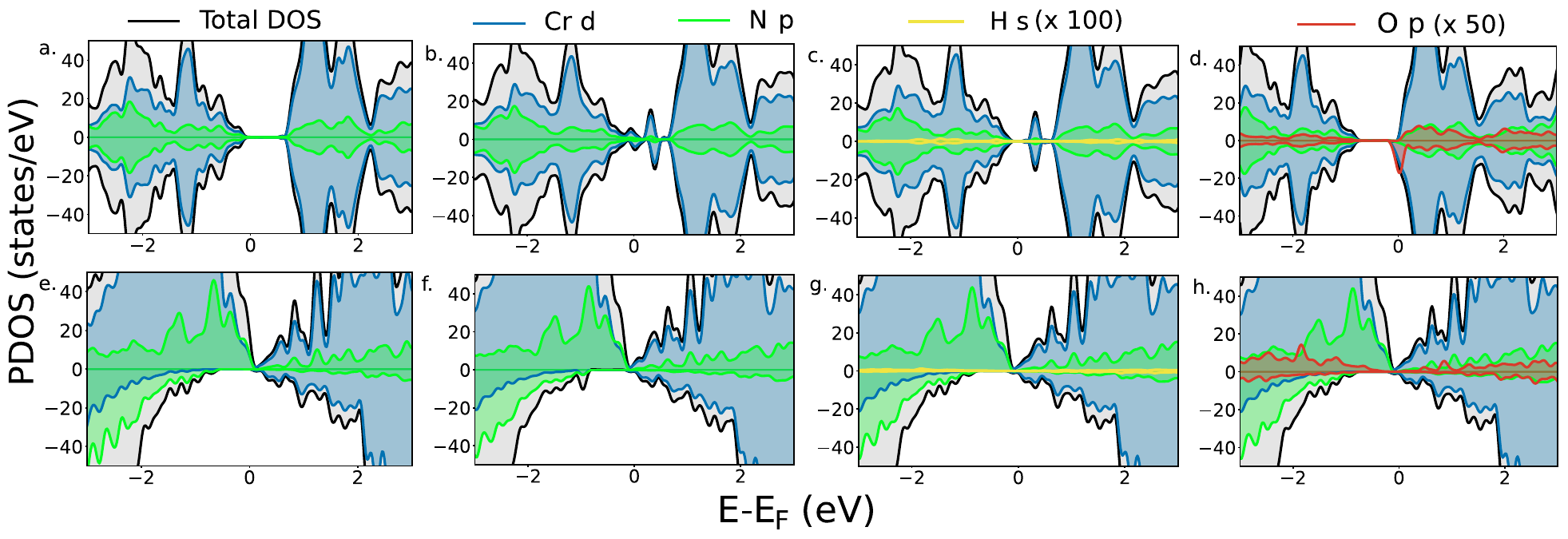}
        \caption{\label{fig4} Projected density of states of (a) pristine (b) $V_{\text{N}}^\times$ (c) $\text{H}_{\text{N}}^\times$ (d) $\text{O}_{\text{N}}^\times$ in CrN\textsubscript{2} and (e) pristine (f) $V_{\text{N}}^\times$ (g) $\text{H}_{i}^\times$ (h) $\text{O}_{\text{N}}^\times$ in CrN.
        }
\end{figure*}

Having established the stability of the most energetically probable defects under equilibrium conditions, we next examine their influence on the electronic structure of CrN\textsubscript{2} and CrN, through projected density of states (PDOS) plots (Figure~\ref{fig4}). Additionally, Bader charge analysis ( Table~\ref{table2}), is employed to quantify local redistribution of electron density around the defect sites and to investigate charge transfer and bonding effects. The range of charges for the whole system is included in Table S4 in the Supplementary Information. CrN\textsubscript{2} in its pristine form (Figure~\ref{fig4}a) exhibits a spin-symmetric landscape with a calculated band gap of 0.73 eV which falls within the range of previously calculated theoretical works (between 0.5 and 1.14 eV)\cite{Zhao2016PotentiallyPressure,Liu2019TheStudy}. By contrast, the PDOS of CrN, shown in Figure~\ref{fig4}(e), reflects a ferromagnetic ordering, with distinct spin-up and spin-down channels. The spin-up channel exhibits metallic behaviour, while the spin-down channel shows a partial gap, giving rise to half-metallicity. This spin-resolved asymmetry originates from exchange splitting of the Cr $d$-states. Orbital-resolved analysis reveals that the occupied states near the Fermi level primarily originate from the Cr \textit{$t_{2g}$} orbitals (particularly \textit{$d_{xz}$}, Figure S4b in Supplementary Information) while the \textit{$e_g$} orbitals lie higher in energy and are largely unoccupied. Bader charge analysis reveals a charge transfer of 1.39e from each Cr atom in CrN\textsubscript{2}, distributed across two neighbouring N atoms, indicating mixed ionic-covalent bonding. In CrN, a slightly higher transfer of 1.43e occurs to one N, reflecting more ionic character.

\begin{table}[h]
    \caption{Bader charges (q\textsubscript{i}, i=Cr, N, O, H) for the pristine and defective compositions of Cr\textsubscript{x}N\textsubscript{y}. q\textsubscript{i} for the defective structures refers to the the nearest neighbours species around the defect site of the lowest energy neutral defects. The full range of Bader charges is provided in Table S4 in the Supplementary Information }
\label{table2} 
\begin{tabular*}{\columnwidth}{@{\extracolsep{\fill}}lllllll}

\hline
System & q\textsubscript{Cr} (e) & q\textsubscript{N} (e) & q\textsubscript{O,H} (e)\\
\hline
\textbf{CrN\textsubscript{2}} &  &   \\
Pristine & 1.39 & -0.63, -0.76 & &  \\
$V_{\text{N}}^\times$ & 1.30 & -1.04 & \\
$\text{H}_{\text{N}}^\times$ & 1.25 & -1.09 & 0.24 \\
$\text{O}_{\text{N}}^\times$ & 1.40 & -0.53 & -0.79 \\
\textbf{CrN} &   &  & \\
Pristine &  1.43 &  -1.43 &\\
$V_{\text{N}}^\times$ & 1.27  & -1.46  &   \\
$\text{H}_{i}^\times$ & 1.54  & -1.50  & 0.27\\
$\text{O}_{\text{N}}^\times$ & 1.50 & -1.45 & -1.39\\
\hline

\end{tabular*}
\end{table}

In \ce{CrN2}, the introduction of a nitrogen vacancy ($V_{\text{N}}^\times$) leads to under-coordination of adjacent Cr atoms, resulting in the formation of localized states within the band gap, primarily consisting of Cr $d$-orbitals with minor N $p$-character (Figure~\ref{fig4}(b)). These states are strongly spin-polarized, reflecting broken spin symmetry due to electron localization. Bader charge analysis supports this, revealing local charge accumulation on the Cr atoms ($q_{\text{Cr}} = 1.30e$) surrounding the vacancy. The neighbouring N atom exhibits a more negative Bader charge ($q_{\text{N}} = -1.04e$), indicative of partial retention of charge from the disrupted N-N dimer. The unpaired electron localizes primarily on Cr sites, reinforcing the observed $d$-dominated in-gap states. The emergence of distinct charge states and their energetic positions are described in the Supplementary Information (Table S3). In contrast to the localized and spin-asymmetric response in CrN\textsubscript{2}, $V_{\text{N}}^\times$ in CrN leads to a redistribution of electronic states around the Fermi level without opening a gap. This primarily affects the spin-down channel, where N $p$ and Cr $d$ states are strongly hybridized (Figure~\ref{fig4}(f)). A reduction in the Bader charges on nearby Cr atoms from 1.43e to 1.27e is seen, consistent with redistribution of electron density toward these sites. N atoms largely retain their original charge (1.43e), with slight increases to 1.46e observed on atoms further from the defect. Despite this local reorganization, the overall metallic character is preserved. This suggests that the defect is efficiently screened, highlighting CrN’s more delocalized, metallic bonding framework and its relative electronic rigidity compared to CrN\textsubscript{2}.

The ability of a hydrogen atom to alter the band gaps of semiconducting nitrides has been observed in several works\ \cite{Pankove1991HydrogenSemiconductors,Pearton1992HydrogenSemiconductors,Pettinari2014HydrogenNanostructuring}, and the ease with which hydrogen can incorporate in a material makes it a common impurity. In \ce{CrN2}, $\text{H}_{\text{N}}^\times$ introduces shallow occupied states near the conduction band edge (Figure~\ref{fig4}(c)), accompanied by a slight upward shift in the Fermi level. The Bader charge value on H is 0.24e, reflecting partial electron donation to nearby Cr atoms that accommodate the excess charge ($q_{\text{Cr}} = 1.25e$). No deep in-gap states are observed, indicating that $\text{H}_{\text{N}}^\times$ behaves as a shallow donor in its neutral state. The minimal spectral weight of H $s$-orbitals suggests weak hybridization and reduced bonding involvement compared to nitrogen, implying that Cr-H interactions are more ionic in nature. This points to hydrogen substitution as a means of tuning carrier concentrations through Fermi level modulation in \ce{CrN2}. 
In contrast, in CrN the $\text{H}_{\text{i}}^\times$ exhibits different behaviour. CrN’s metallic nature and high carrier density result in efficient screening of the defect, with no emergence of discrete defect levels within the band structure (Figure~\ref{fig4}(g)). The Fermi level remains pinned, and the overall PDOS shape shows minimal distortion apart from a modest increase in Cr 3$d$ state intensity just below the Fermi level. This aligns with the Bader charge on H (0.27e), indicating partial electron donation. The nearest neighbour Cr atoms become more oxidized ($q_{\text{Cr}} = 1.54e$), while N atoms also exhibit increased negative charge ($q_{\text{N}} = -1.50e$) compared to the pristine case, suggesting a localized charge redistribution in response to the interstitial. Unlike in \ce{CrN2}, where H replaces a more electronegative N atom, the interstitial configuration in CrN leads to more delocalized behaviour with limited impact on the electronic structure. While $\text{H}_{\text{N}}^\times$ in \ce{CrN2} acts as a shallow donor, $\text{H}_{\text{i}}^\times$ in CrN perturbs the system only weakly, with limited consequences for carrier trapping or Fermi level control.

The electronic structure of CrN\textsubscript{2} and CrN in the presence of an oxygen atom substituted on a nitrogen site ($\text{O}_{\text{N}}^{\times}$) is shown in Figure~\ref{fig4}(d) and Figure~\ref{fig4}(h), respectively. $\text{O}_{\text{N}}^{\times}$ in CrN\textsubscript{2}, leads to a small increase in electron accumulation at the oxygen site (-0.79e). This is accompanied by a redistribution of charge within the local environment where the nearest neighbouring Cr atom becomes sligtly more oxidised ($+0.10e$) 
and the neighbouring N becomes less negative by a comparable amount (+0.10e), indicating a broader shift in electron density beyond the immediate substitution site. The PDOS reveals hybridisation of Cr $d$-states and O $p$-orbitals in the spin-down channel near the conduction band, alongside a clear shift of the Fermi level toward the conduction band edge. This defect stabilizes a +1 charge state ($\text{O}_{\text{N}}^{\bullet}$), as observed in Figure S8 in the Supplementary Information. In CrN, $\text{O}_{\text{N}}^{\times}$ enhances local electron withdrawal due to oxygen’s higher electronegativity. This is reflected in the $0.04e$ change in Bader charge, with oxygen carrying $-1.39e$, compared to $-1.43e$ for nitrogen in the pristine lattice. As a result, nearby Cr atoms become more oxidised, with the nearest neighbours reaching up to $+1.50e$, compared to the bulk value of $+1.43e$. This confirms localized electron depletion within the Cr-O coordination shell. Nitrogen atoms retain Bader charges close to their pristine values (-1.43e to -1.45e), with only minimal perturbations observed.  Thus, $\text{O}_{\text{N}}^{\times}$ in CrN does influence the local electronic environment, but does not introduce any magnetic disruption, contrasting the spin-polarised behaviour seen in CrN\textsubscript{2}.

Overall, these results highlight the contrasting defect responses of CrN\textsubscript{2} and CrN, closely tied to their defect energetics and magnetic response discussed in the previous section. While defects in CrN\textsubscript{2} tend to introduce localized, spin-polarized states that modulate the electronic structure and magnetic character, CrN exhibits a more delocalized response with efficient screening and preserved metallicity. These differences reflect the underlying bonding environments and electronic rigidity of the two systems, offering insight into how point defects may be leveraged for tuning physical properties in nitride-based materials.

\subsection{Mechanical properties}
With the energetic and electronic behaviour of intrinsic and extrinsic defects established,  their effect on the mechanical properties of CrN\textsubscript{2} and CrN is examined.
This is done in terms of changes in elasticity, plasticity and rigidity relative to the bulk elastic constants, assessing the material's response to mechanical stress, deformation and overall stiffness in the presence of the most energetically stable defects. The derived elastic properties for CrN\textsubscript{2} and CrN are summarized in Table \ref{table7}. The full second-order elastic tensor components (C\textsubscript{ij}) for both pristine and defected structures are provided in Supplementary Information (Subsections 1.1 and 2.1). The Born-Huang criteria confirm that all of the defective structures discussed are mechanically stable. \par

\begin{table*}[htbp]
\small
\caption{Calculated bulk modulus (K\textsubscript{V}) , Young's modulus (E\textsubscript{V}), Shear modulus (G\textsubscript{V}), Vicker's hardness (H\textsubscript{v}), Pugh's ratio (k), Poisson's ratio ($\nu$) and Universal Elastic Anisotropy (A$^U$) of the most stable defects in \ce{CrN2} and CrN.}
\label{table7} 
\begin{tabular*}{\textwidth}{@{\extracolsep{\fill}}lccccccc}
\hline
Crystal Structure & K\textsubscript{V} (GPa) & E\textsubscript{V} (GPa) &  G\textsubscript{V} (GPa) & H\textsubscript{v} (GPa) & k & $\nu$ & A$^U$ \\

\multicolumn{8}{l}{\textbf{\ce{CrN2}}}  \\
Pristine & 337.45 & 593.79 & 246.04 & 31.67 & 1.37 & 0.21 & 0.64  \\ 
$V_{\text{N}}^\times$  & 322.70 & 560.03 & 231.28 & 29.72 & 1.39 & 0.21 & 0.55  \\
$\text{H}_{\text{N}}^\times$ & 329.17 & 571.75 & 236.16 & 30.20 & 1.39 & 0.21 & 0.59  \\
$\text{O}_{\text{N}}^\times$ & 336.36 & 573.61 & 235.90 & 29.41 & 1.43 &  0.22 & 0.63  \\

\multicolumn{8}{l}{\textbf{CrN}} \\

Pristine & 237.76 & 412.87 & 170.53 & 23.98 & 1.40 & 0.21 & 0.05 \\
$V_{\text{N}}^\times$  & 233.97 & 409.55 & 169.48 & 24.14 & 1.39 & 0.21 & 0.05  \\
$\text{H}_{i}^\times$ & 235.82 & 394.09 & 161.32 & 21.84 & 1.46 & 0.22 & 0.02  \\
$\text{O}_{\text{N}}^\times$ & 236.15 & 413.07 & 170.90 & 24.26 & 1.38 & 0.21 & 0.05  \\
\hline

\end{tabular*}

\end{table*}

Pristine CrN\textsubscript{2} exhibits higher elastic moduli than CrN, with average K\textsubscript{V} = 337.45 GPa, E\textsubscript{V} = 593.79 GPa, and G\textsubscript{V} = 246.04, compared to 237.76 GPa, 412.87 GPa and 170.53 GPa for CrN, respectively. These values are consistent with previous studies, where the bulk modulus of \ce{CrN2} has been reported to be around 325 GPa\cite{Liu2019TheStudy} theoretically and 286 GPa\cite{Niwa2019High-pressureDimer} experimentally\cite{Liu2019TheStudy}, while CrN values are 241.33 GPa \cite{Zhou2014StructuralExperiment} theoretically and 260 GPa \cite{Zhang2020MappingScale} experimentally. This difference reflects the directional bonding in hexagonal CrN\textsubscript{2}, particularly the presence of N-N dimers, which provide additional rigidity. In contrast, the cubic symmetry of CrN gives rise to a more isotropic elastic response, consistent with its delocalised, metallic bonding environment. This distinction underpins the differing sensitivities of each material to local perturbations. \par

The introduction of a single $V_{\text{N}}^\times$ leads to changes in the mechanical stiffness of CrN\textsubscript{2} and CrN, with a notably stronger reduction in the former. In CrN\textsubscript{2}, K\textsubscript{V} decreases from 337.45 GPa to 322.70 GPa (-4.37\%), with corresponding decreases in both E\textsubscript{V} from 593.79 GPa to 560.03 GPa (-5.68\%) and G\textsubscript{V} from 246.04 GPa to 231.28 GPa (-5.99\%). This can be traced back to the disruption of its directional N-N bonding network, previously shown (section 3.2) to be sensitive to local charge and spin rearrangements. These reductions are accompanied by changes in directional stiffness but no symmetry breaking evident from Figure S10 and S11 in the Supplementary Information. In comparison, CrN shows less pronounced reductions with K\textsubscript{V} decreasing from 237.76 GPa to 233.97 GPa (-1.6\%). A similar result was reported in  previous work on MoN\textsubscript{x}, where the reduction was calculated to be 2\% \cite{KarimiAghda2023ValenceNitrides}. This trend remains consistent in the case of E\textsubscript{V} and G\textsubscript{V}, where both moduli show a decrease from the pristine value of 412.87 GPa to 409.55 GPa (-0.8\%) and 170.53 GPa to 169.48 GPa (-0.62\%), respectively. Tian’s hardness (H\textsubscript{V}) follows a similar pattern: CrN\textsubscript{2} softens (-6.2\%), while CrN shows a slight increase (+0.7\%), aligning more closely with experimental values \cite{Wang2016SynthesisCrN}. 
CrN thus remains largely isotropic in mechanical response. This indicates the intrinsic presence of such defects in CrN, consistent with its low defect formation energy (Table~\ref{table1}). These results show that CrN\textsubscript{2}, with its more anisotropic bonding structure, is more susceptible to defect-induced mechanical degradation than CrN, which retains its stiffness and isotropy in the presence of $V_{\text{N}}^\times$. \par


The mechanical response of CrN\textsubscript{2} and CrN to hydrogen-related point defects reveals distinct trends in both isotropic and directional hardness. In CrN\textsubscript{2}, $\text{H}_{\text{N}}^\times$ leads to moderate reductions in K\textsubscript{V} from 337.45 to 329.17 GPa (-2.5\%), E\textsubscript{V} from 593.79 to 571.75 GPa (-3.7\%), and G\textsubscript{V} from 246.04 to 236.16 GPa (-4.0\%). H\textsubscript{V} decreases by 4.7\% (from 31.67 to 30.20 GPa), while Pugh’s ratio slightly increases (from 1.37 to 1.39), and A$^U$ drops from 0.64 to 0.59. This is linked to local lattice expansion and weakened Cr-H bonding, as suggested by the structural relaxation and partial charge transfer in Sections 3.1 and 3.2. Directional E\textsubscript{V} shown in Figure S10 in the Supplementary Information indicate uniform but mild softening across crystallographic planes, without significant symmetry disruption. In CrN, $\text{H}_{i}^\times$ reduces K\textsubscript{V} from 237.76 to 235.82 (-0.82\%), E\textsubscript{V} from 412.87 to 394.09 GPa (-4.55\%), G\textsubscript{V} from 170.53 to 161.32 GPa (-5.4\%), and H\textsubscript{V} from 23.98 to 21.84 GPa (-8.9\%). While A$^U$ drops from 0.05 to 0.02, indicating an overall isotropic trend, directional modulus plots (Figure S11) reveal non-trivial changes in elastic symmetry. In the (001) plane, the [001] direction transitions from a faceted to a nearly circular profile, while [100] and [010] retain their symmetry, suggesting reduced anisotropy along the out-of-plane axis and in-plane lattice distortion. In the (011) plane (Figure S11) , the [0$\overline{1}$1] direction becomes more circular, while [100] shows deviation, pointing to preferential elastic relaxation along diagonally aligned bonds. The (111) plane (Figure S11) exhibits unique directional profiles, with [1$\overline{1}$0] softening more prominently than [0$\overline{1}$1] or [111], indicating local symmetry breaking. These results highlight how $\text{H}_{i}^\times$ induces anisotropic elastic softening in CrN, with varying directional effects depending on crystallographic orientation. Thus, CrN\textsubscript{2} shows modest and mostly isotropic stiffness reductions whereas CrN displays direction-dependent distortions and subtle symmetry loss especially in high-index planes, reflecting the sensitivity of its cubic lattice to local bonding perturbations. This stronger response in CrN can be attributed to the interstitial incorporation of hydrogen, which disrupts its delocalised metallic bonding network and induces local anisotropic lattice distortions. In contrast, CrN\textsubscript{2}, accommodates hydrogen at a substitutional site, resulting in more localised and symmetric relaxations. These directionally selective distortions may contribute to localized stress accumulation, providing a mechanistic link to hydrogen-induced embrittlement in CrN-based hard coatings under mechanical loading. \par
The presence of a single $\text{O}_{\text{N}}^\times$ in CrN\textsubscript{2} reduces K\textsubscript{V} from 337.45 to 336.36 GPa (-0.36\%), E\textsubscript{V} from 593.79 to 573.61 GPa (-3.42\%) and G\textsubscript{V} from 246.04 to 235.90 GPa (-4.15\%). A more pronounced drop is observed in H\textsubscript{V}, which falls by over 7\% from 31.67 to 29.41 GPa, accompanied by a slight decrease in A\textsuperscript{U} from 0.64 to 0.63. These changes reflect a local weakening of the Cr-O bonding environment, consistent with oxygen’s enhanced electronegativity and charge accumulation observed in the earlier electronic analysis (section 3.2). Despite these macroscopic effects, directional E\textsubscript{V} plots remain nearly unchanged across most planes. A subtle asymmetry appears in the (101) plane, where the [101] and [$\overline{1}$01] directions no longer exhibit identical stiffness profiles, suggesting that oxygen induces weak but localised anisotropic distortions. In CrN, the mechanical response of $\text{O}_{\text{N}}^\times$ is more subdued. K\textsubscript{V} decreases from 237.76 to 236.15 GPa (-0.68\%,) while both E\textsubscript{V} and G\textsubscript{V} increase by 0.04\% and 0.2\% respectively. H\textsubscript{V} increases by 1.17\%, implying enhanced resistance to plastic deformation. This aligns with experimental observations of improved hardness of chromium oxynitride as compared to CrN \cite{Du2021InfluenceCoating}. Directional E\textsubscript{V} plots show no deviation from the pristine structure across all crystallographic planes, indicating that oxygen substitution induces minimal elastic perturbation in CrN’s cubic lattice. \par 
Thus, CrN\textsubscript{2}, despite its higher intrinsic directional hardness than CrN, exhibits greater sensitivity to impurity-induced softening particularly due to the disruption of directional N-N bonding. In contrast, $V_{\text{N}}^\times$ and $\text{O}_{\text{N}}^\times$ cause minor changes in CrN's hardness with the latter slightly improving it. The introduction of hydrogen, which had not previously been examined in this context, results in more significant softening in CrN than in CrN\textsubscript{2}, and introduces directional distortions linked to local bonding rearrangements. These findings suggest that while CrN\textsubscript{2} may show superior initial mechanical metrics, CrN is likely to be more robust under non-ideal processing conditions or extended operational use where point defects and impurities are unavoidable.

\section{Conclusions}
Using density functional theory simulations, we systematically investigated the structural, electronic, and mechanical effects of intrinsic and extrinsic point defects in two hard Cr-N phases: cubic CrN and hexagonal CrN\textsubscript{2}. By systematically analysing defect formation energetics, local structural distortions, magnetic responses, and elastic behaviour, this study highlights the contrasting defect tolerance of the two materials and the underlying bonding mechanisms that govern these responses. \par  
Among intrinsic defects, $V_{\text{N}}^\times$ is the most energetically favourable defect in both the compositions. However, its impact on electronic and mechanical properties is markedly different: in CrN\textsubscript{2}, $V_{\text{N}}^\times$ localises charge and spin, introducing states within the band gap and modifying magnetic character. It causes significant reductions in all elastic moduli and hardness, accompanied by directional softening and contour shrinkage in the E\textsubscript{V} plots. In contrast, CrN retains its magnetic character and near-isotropic stiffness, with negligible directional changes, demonstrating its higher tolerance to such defects. Extrinsic impurities show greater variation. In CrN\textsubscript{2}, hydrogen and oxygen preferentially substitute at nitrogen sites, both leading to modified occupation of electron density and reductions in stiffness, with $\text{O}_{\text{N}}^\times$ resulting in the largest hardness drop (7.1\%). These effects are mostly isotropic, though subtle anisotropic softening appears in certain pyramidal planes. In CrN, hydrogen stabilizes as an interstitial and induces moderate mechanical softening (hardness decreases by over 8\%), accompanied by visible symmetry breaking in directional modulus plots, particularly in high-index planes. Oxygen, however, substitutes nitrogen and causes no major change in elasticity or anisotropy, while slightly increasing hardness.  These results reveal that CrN\textsubscript{2}, despite its higher intrinsic hardness, is more sensitive to point defect-induced softening, particularly due to its anisotropic N-N bonding network. CrN, with its highly symmetric Cr-N framework, demonstrates greater defect tolerance but remains vulnerable to hydrogen-induced symmetry breaking and softening, relevant to its known susceptibility to hydrogen embrittlement. This work provides fundamental insights into the defect landscapes of Cr-N systems, and demonstrates how point defects can directionally modulate the mechanical performance of Cr-based nitrides. While CrN offers superior defect tolerance, CrN\textsubscript{2} exhibits higher hardness but greater sensitivity to defect-induced softening. These insights provide a pathway for defect-engineered design of durable, high-performance hard coatings in reactive or hydrogen-rich environments.

\section*{Conflicts of interest}
The authors declare no conflicts of interest.

\section*{Data availability}
The data supporting this article have been included as part of the Supplementary Information.

\section{Acknowledgements}
This work was conducted at the Advanced Research Center for Nanolithography, a public–private partnership between the University of Amsterdam (UvA), Vrije Universiteit Amsterdam (VU), Rijksuniversiteit Groningen (RUG), the Netherlands Organization for Scientific Research (NWO), and the semiconductor equipment manufacturer ASML. This work made use of the Dutch national e-infrastructure with the support of the SURF Cooperative using grant no. EINF-5669 and EINF-9637 and the authors thank SURF (www.surf.nl) for the support in using the National Supercomputer Snellius. E.O. is grateful for a WISE Fellowship from the NWO and support via Holland High Tech through a public-private partnership in research and development within the Dutch top sector of High-Tech Systems and Materials (HTSM).

\bibliography{crn}

\providecommand*{\mcitethebibliography}{\thebibliography}
\csname @ifundefined\endcsname{endmcitethebibliography}
{\let\endmcitethebibliography\endthebibliography}{}
\begin{mcitethebibliography}{103}
\providecommand*{\natexlab}[1]{#1}
\providecommand*{\mciteSetBstSublistMode}[1]{}
\providecommand*{\mciteSetBstMaxWidthForm}[2]{}
\providecommand*{\mciteBstWouldAddEndPuncttrue}
  {\def\EndOfBibitem{\unskip.}}
\providecommand*{\mciteBstWouldAddEndPunctfalse}
  {\let\EndOfBibitem\relax}
\providecommand*{\mciteSetBstMidEndSepPunct}[3]{}
\providecommand*{\mciteSetBstSublistLabelBeginEnd}[3]{}
\providecommand*{\EndOfBibitem}{}
\mciteSetBstSublistMode{f}
\mciteSetBstMaxWidthForm{subitem}
{(\emph{\alph{mcitesubitemcount}})}
\mciteSetBstSublistLabelBeginEnd{\mcitemaxwidthsubitemform\space}
{\relax}{\relax}

\bibitem[Hones \emph{et~al.}(1997)Hones, Sanjines, and Levy]{Hones1997CharacterizationCoatings}
P.~Hones, R.~Sanjines and F.~Levy, \emph{Surface and Coatings Technology}, 1997, \textbf{94-95}, 398--402\relax
\mciteBstWouldAddEndPuncttrue
\mciteSetBstMidEndSepPunct{\mcitedefaultmidpunct}
{\mcitedefaultendpunct}{\mcitedefaultseppunct}\relax
\EndOfBibitem
\bibitem[Wuhrer and Yeung(2004)]{Wuhrer2004ACoatings}
R.~Wuhrer and W.~Y. Yeung, \emph{Scripta Materialia}, 2004, \textbf{50}, 1461--1466\relax
\mciteBstWouldAddEndPuncttrue
\mciteSetBstMidEndSepPunct{\mcitedefaultmidpunct}
{\mcitedefaultendpunct}{\mcitedefaultseppunct}\relax
\EndOfBibitem
\bibitem[Fuentes \emph{et~al.}(2005)Fuentes, Rodriguez, Avelar-Batista, Housden, Montal{\'{a}}, Carreras, Crist{\'{o}}bal, Damborenea, and Tate]{Fuentes2005RecentProtection}
G.~Fuentes, R.~Rodriguez, J.~Avelar-Batista, J.~Housden, F.~Montal{\'{a}}, L.~Carreras, A.~Crist{\'{o}}bal, J.~Damborenea and T.~Tate, \emph{Journal of Materials Processing Technology}, 2005, \textbf{167}, 415--421\relax
\mciteBstWouldAddEndPuncttrue
\mciteSetBstMidEndSepPunct{\mcitedefaultmidpunct}
{\mcitedefaultendpunct}{\mcitedefaultseppunct}\relax
\EndOfBibitem
\bibitem[Holleck and Schier(1995)]{Holleck1995MultilayerProtection}
H.~Holleck and V.~Schier, \emph{Surface and Coatings Technology}, 1995, \textbf{76-77}, 328--336\relax
\mciteBstWouldAddEndPuncttrue
\mciteSetBstMidEndSepPunct{\mcitedefaultmidpunct}
{\mcitedefaultendpunct}{\mcitedefaultseppunct}\relax
\EndOfBibitem
\bibitem[Senf and Broszeit(1999)]{Senf1999WearCoatings}
J.~Senf and E.~Broszeit, \emph{Advanced Engineering Materials}, 1999, \textbf{1}, 133--137\relax
\mciteBstWouldAddEndPuncttrue
\mciteSetBstMidEndSepPunct{\mcitedefaultmidpunct}
{\mcitedefaultendpunct}{\mcitedefaultseppunct}\relax
\EndOfBibitem
\bibitem[Jagielski \emph{et~al.}(2000)Jagielski, Khanna, Kucinski, Mishra, Racolta, Sioshansi, Tobin, Thereska, Uglov, Vilaithong, Viviente, Yang, and Zalar]{Jagielski2000EffectSteel}
J.~Jagielski, A.~S. Khanna, J.~Kucinski, D.~S. Mishra, P.~Racolta, P.~Sioshansi, E.~Tobin, J.~Thereska, V.~Uglov, T.~Vilaithong, J.~Viviente, S.~Z. Yang and A.~Zalar, \emph{Applied Surface Science}, 2000, \textbf{156}, 47--64\relax
\mciteBstWouldAddEndPuncttrue
\mciteSetBstMidEndSepPunct{\mcitedefaultmidpunct}
{\mcitedefaultendpunct}{\mcitedefaultseppunct}\relax
\EndOfBibitem
\bibitem[Bertrand \emph{et~al.}(2000)Bertrand, Mahdjoub, and Meunier]{Bertrand2000ACoatings}
G.~Bertrand, H.~Mahdjoub and C.~Meunier, \emph{Surface and Coatings Technology}, 2000, \textbf{126}, 199--209\relax
\mciteBstWouldAddEndPuncttrue
\mciteSetBstMidEndSepPunct{\mcitedefaultmidpunct}
{\mcitedefaultendpunct}{\mcitedefaultseppunct}\relax
\EndOfBibitem
\bibitem[Schauperl \emph{et~al.}(2008)Schauperl, Ivu{\v{s}}i{\'{c}}, and Runje]{Schauperl2008WearCoatings}
Z.~Schauperl, V.~Ivu{\v{s}}i{\'{c}} and B.~Runje, \emph{Materials Testing}, 2008, \textbf{50}, 326--331\relax
\mciteBstWouldAddEndPuncttrue
\mciteSetBstMidEndSepPunct{\mcitedefaultmidpunct}
{\mcitedefaultendpunct}{\mcitedefaultseppunct}\relax
\EndOfBibitem
\bibitem[Evseev \emph{et~al.}(2020)Evseev, Ziatdinov, and Tolynbekov]{Evseev2020ChromiumMode}
N.~S. Evseev, M.~K. Ziatdinov and A.~B. Tolynbekov, \emph{Journal of Physics: Conference Series}, 2020, \textbf{1611}, 012036\relax
\mciteBstWouldAddEndPuncttrue
\mciteSetBstMidEndSepPunct{\mcitedefaultmidpunct}
{\mcitedefaultendpunct}{\mcitedefaultseppunct}\relax
\EndOfBibitem
\bibitem[Hultman(2000)]{Hultman2000ThermalFilms}
L.~Hultman, \emph{Vacuum}, 2000, \textbf{57}, 1--30\relax
\mciteBstWouldAddEndPuncttrue
\mciteSetBstMidEndSepPunct{\mcitedefaultmidpunct}
{\mcitedefaultendpunct}{\mcitedefaultseppunct}\relax
\EndOfBibitem
\bibitem[Eklund \emph{et~al.}(2016)Eklund, Kerdsongpanya, and Alling]{Eklund2016Transition-metal-nitride-basedMaterials}
P.~Eklund, S.~Kerdsongpanya and B.~Alling, \emph{Journal of Materials Chemistry C}, 2016, \textbf{4}, 3905--3914\relax
\mciteBstWouldAddEndPuncttrue
\mciteSetBstMidEndSepPunct{\mcitedefaultmidpunct}
{\mcitedefaultendpunct}{\mcitedefaultseppunct}\relax
\EndOfBibitem
\bibitem[Esaka \emph{et~al.}(1997)Esaka, Furuya, Shimada, Imamura, Matsubayashi, Sato, Nishijima, Kawana, Ichimura, and Kikuchi]{Esaka1997ComparisonSpectroscopy}
F.~Esaka, K.~Furuya, H.~Shimada, M.~Imamura, N.~Matsubayashi, H.~Sato, A.~Nishijima, A.~Kawana, H.~Ichimura and T.~Kikuchi, \emph{Journal of Vacuum Science {\&} Technology A: Vacuum, Surfaces, and Films}, 1997, \textbf{15}, 2521--2528\relax
\mciteBstWouldAddEndPuncttrue
\mciteSetBstMidEndSepPunct{\mcitedefaultmidpunct}
{\mcitedefaultendpunct}{\mcitedefaultseppunct}\relax
\EndOfBibitem
\bibitem[Chim \emph{et~al.}(2009)Chim, Ding, Zeng, and Zhang]{Chim2009OxidationArc}
Y.~Chim, X.~Ding, X.~Zeng and S.~Zhang, \emph{Thin Solid Films}, 2009, \textbf{517}, 4845--4849\relax
\mciteBstWouldAddEndPuncttrue
\mciteSetBstMidEndSepPunct{\mcitedefaultmidpunct}
{\mcitedefaultendpunct}{\mcitedefaultseppunct}\relax
\EndOfBibitem
\bibitem[Vera \emph{et~al.}(2011)Vera, Vite, Lewis, Gallardo, and Laguna-Camacho]{Vera2011ASubstrates}
E.~Vera, M.~Vite, R.~Lewis, E.~Gallardo and J.~Laguna-Camacho, \emph{Wear}, 2011, \textbf{271}, 2116--2124\relax
\mciteBstWouldAddEndPuncttrue
\mciteSetBstMidEndSepPunct{\mcitedefaultmidpunct}
{\mcitedefaultendpunct}{\mcitedefaultseppunct}\relax
\EndOfBibitem
\bibitem[Qi \emph{et~al.}(2013)Qi, Liu, Wu, Zhu, Wang, and Wu]{Qi2013ACoatings}
Z.~Qi, B.~Liu, Z.~Wu, F.~Zhu, Z.~Wang and C.~Wu, \emph{Thin Solid Films}, 2013, \textbf{544}, 515--520\relax
\mciteBstWouldAddEndPuncttrue
\mciteSetBstMidEndSepPunct{\mcitedefaultmidpunct}
{\mcitedefaultendpunct}{\mcitedefaultseppunct}\relax
\EndOfBibitem
\bibitem[Du \emph{et~al.}(2020)Du, Chen, Chen, and Du]{Du2020MechanicalCoatings}
J.~W. Du, L.~Chen, J.~Chen and Y.~Du, \emph{Vacuum}, 2020, \textbf{179}, 109468\relax
\mciteBstWouldAddEndPuncttrue
\mciteSetBstMidEndSepPunct{\mcitedefaultmidpunct}
{\mcitedefaultendpunct}{\mcitedefaultseppunct}\relax
\EndOfBibitem
\bibitem[Rahman and Atta~Ogwu(2020)]{Rahman2020CorrosionApplications}
S.~U. Rahman and A.~Atta~Ogwu, \emph{Advances in Medical and Surgical Engineering}, 2020,  239--265\relax
\mciteBstWouldAddEndPuncttrue
\mciteSetBstMidEndSepPunct{\mcitedefaultmidpunct}
{\mcitedefaultendpunct}{\mcitedefaultseppunct}\relax
\EndOfBibitem
\bibitem[Santhanam(1996)]{Santhanam1996ApplicationTools}
A.~T. Santhanam, \emph{The Chemistry of Transition Metal Carbides and Nitrides}, 1996,  28--52\relax
\mciteBstWouldAddEndPuncttrue
\mciteSetBstMidEndSepPunct{\mcitedefaultmidpunct}
{\mcitedefaultendpunct}{\mcitedefaultseppunct}\relax
\EndOfBibitem
\bibitem[Shabashov \emph{et~al.}(2022)Shabashov, Lyashkov, Zamatovskii, Kozlov, Kataeva, Novikov, and Ustyugov]{Shabashov2022MechanosynthesisNitrides}
V.~Shabashov, K.~Lyashkov, A.~Zamatovskii, K.~Kozlov, N.~Kataeva, E.~Novikov and Y.~Ustyugov, \emph{Materials 2022, Vol. 15, Page 5038}, 2022, \textbf{15}, 5038\relax
\mciteBstWouldAddEndPuncttrue
\mciteSetBstMidEndSepPunct{\mcitedefaultmidpunct}
{\mcitedefaultendpunct}{\mcitedefaultseppunct}\relax
\EndOfBibitem
\bibitem[Singh \emph{et~al.}(2020)Singh, Chattopadhyaya, Pramanik, Kumar, Basak, Pandey, Murtaza, Legutko, and Litak]{Singh2020TribologicalTemperature}
S.~K. Singh, S.~Chattopadhyaya, A.~Pramanik, S.~Kumar, A.~K. Basak, S.~M. Pandey, Q.~Murtaza, S.~Legutko and G.~Litak, \emph{Materials}, 2020, \textbf{13}, 4497\relax
\mciteBstWouldAddEndPuncttrue
\mciteSetBstMidEndSepPunct{\mcitedefaultmidpunct}
{\mcitedefaultendpunct}{\mcitedefaultseppunct}\relax
\EndOfBibitem
\bibitem[Jiang \emph{et~al.}(1993)Jiang, Goto, and Hirai]{Jiang1993Non-stoichiometryDeposition}
C.-C. Jiang, T.~Goto and T.~Hirai, \emph{Journal of Alloys and Compounds}, 1993, \textbf{190}, 197--200\relax
\mciteBstWouldAddEndPuncttrue
\mciteSetBstMidEndSepPunct{\mcitedefaultmidpunct}
{\mcitedefaultendpunct}{\mcitedefaultseppunct}\relax
\EndOfBibitem
\bibitem[Shin \emph{et~al.}(2003)Shin, Gall, Hellgren, Patscheider, Petrov, and Greene]{Shin2003VacancyLayers}
C.-S. Shin, D.~Gall, N.~Hellgren, J.~Patscheider, I.~Petrov and J.~E. Greene, \emph{Journal of Applied Physics}, 2003, \textbf{93}, 6025--6028\relax
\mciteBstWouldAddEndPuncttrue
\mciteSetBstMidEndSepPunct{\mcitedefaultmidpunct}
{\mcitedefaultendpunct}{\mcitedefaultseppunct}\relax
\EndOfBibitem
\bibitem[Shin \emph{et~al.}(2004)Shin, Rudenja, Gall, Hellgren, Lee, Petrov, and Greene]{Shin2004GrowthMgO001}
C.-S. Shin, S.~Rudenja, D.~Gall, N.~Hellgren, T.-Y. Lee, I.~Petrov and J.~E. Greene, \emph{Journal of Applied Physics}, 2004, \textbf{95}, 356--362\relax
\mciteBstWouldAddEndPuncttrue
\mciteSetBstMidEndSepPunct{\mcitedefaultmidpunct}
{\mcitedefaultendpunct}{\mcitedefaultseppunct}\relax
\EndOfBibitem
\bibitem[Vissers \emph{et~al.}(2013)Vissers, Gao, Kline, Sandberg, Weides, Wisbey, and Pappas]{Vissers2013CharacterizationGrowth}
M.~R. Vissers, J.~Gao, J.~S. Kline, M.~Sandberg, M.~P. Weides, D.~S. Wisbey and D.~P. Pappas, \emph{Thin Solid Films}, 2013, \textbf{548}, 485--488\relax
\mciteBstWouldAddEndPuncttrue
\mciteSetBstMidEndSepPunct{\mcitedefaultmidpunct}
{\mcitedefaultendpunct}{\mcitedefaultseppunct}\relax
\EndOfBibitem
\bibitem[Skala and Capkova(1990)]{Skala1990NitrogenNitride}
L.~Skala and P.~Capkova, \emph{Journal of Physics: Condensed Matter}, 1990, \textbf{2}, 8293--8301\relax
\mciteBstWouldAddEndPuncttrue
\mciteSetBstMidEndSepPunct{\mcitedefaultmidpunct}
{\mcitedefaultendpunct}{\mcitedefaultseppunct}\relax
\EndOfBibitem
\bibitem[Gueddaoui \emph{et~al.}(2006)Gueddaoui, Schmerber, Abes, Guemmaz, and Parlebas]{Gueddaoui2006EffectsFilms}
H.~Gueddaoui, G.~Schmerber, M.~Abes, M.~Guemmaz and J.~Parlebas, \emph{Catalysis Today}, 2006, \textbf{113}, 270--274\relax
\mciteBstWouldAddEndPuncttrue
\mciteSetBstMidEndSepPunct{\mcitedefaultmidpunct}
{\mcitedefaultendpunct}{\mcitedefaultseppunct}\relax
\EndOfBibitem
\bibitem[Gall \emph{et~al.}(2002)Gall, Shin, Spila, Od{\'{e}}n, Senna, Greene, and Petrov]{Gall2002GrowthProperties}
D.~Gall, C.~S. Shin, T.~Spila, M.~Od{\'{e}}n, M.~J. Senna, J.~E. Greene and I.~Petrov, \emph{Journal of Applied Physics}, 2002, \textbf{91}, 3589\relax
\mciteBstWouldAddEndPuncttrue
\mciteSetBstMidEndSepPunct{\mcitedefaultmidpunct}
{\mcitedefaultendpunct}{\mcitedefaultseppunct}\relax
\EndOfBibitem
\bibitem[Hones \emph{et~al.}(2003)Hones, Martin, Regula, and L{\'{e}}vy]{Hones2003StructuralFilms}
P.~Hones, N.~Martin, M.~Regula and F.~L{\'{e}}vy, \emph{J. Phys. D: Appl. Phys}, 2003, \textbf{36}, 1023--1029\relax
\mciteBstWouldAddEndPuncttrue
\mciteSetBstMidEndSepPunct{\mcitedefaultmidpunct}
{\mcitedefaultendpunct}{\mcitedefaultseppunct}\relax
\EndOfBibitem
\bibitem[Subramanian \emph{et~al.}(2012)Subramanian, Prabakaran, and Jayachandran]{Subramanian2012InfluenceSputtering}
B.~Subramanian, K.~Prabakaran and M.~Jayachandran, \emph{Bulletin of Materials Science}, 2012, \textbf{35}, 505--511\relax
\mciteBstWouldAddEndPuncttrue
\mciteSetBstMidEndSepPunct{\mcitedefaultmidpunct}
{\mcitedefaultendpunct}{\mcitedefaultseppunct}\relax
\EndOfBibitem
\bibitem[Zhang \emph{et~al.}(2013)Zhang, Li, Daniel, Mitterer, and Dehm]{Zhang2013InsightsCrN}
Z.~Zhang, H.~Li, R.~Daniel, C.~Mitterer and G.~Dehm, \emph{Physical Review B - Condensed Matter and Materials Physics}, 2013, \textbf{87}, 014104\relax
\mciteBstWouldAddEndPuncttrue
\mciteSetBstMidEndSepPunct{\mcitedefaultmidpunct}
{\mcitedefaultendpunct}{\mcitedefaultseppunct}\relax
\EndOfBibitem
\bibitem[Gregoryanz \emph{et~al.}(2004)Gregoryanz, Sanloup, Somayazulu, Badro, Fiquet, Mao, and Hemley]{Gregoryanz2004SynthesisNitride}
E.~Gregoryanz, C.~Sanloup, M.~Somayazulu, J.~Badro, G.~Fiquet, H.-k. Mao and R.~J. Hemley, \emph{Nature Materials}, 2004, \textbf{3}, 294--297\relax
\mciteBstWouldAddEndPuncttrue
\mciteSetBstMidEndSepPunct{\mcitedefaultmidpunct}
{\mcitedefaultendpunct}{\mcitedefaultseppunct}\relax
\EndOfBibitem
\bibitem[Niwa \emph{et~al.}(2014)Niwa, Suzuki, Muto, Tatsumi, Soda, Kikegawa, and Hasegawa]{Niwa2014Discoverysub2/sub}
K.~Niwa, K.~Suzuki, S.~Muto, K.~Tatsumi, K.~Soda, T.~Kikegawa and M.~Hasegawa, \emph{Chemistry – A European Journal}, 2014, \textbf{20}, 13885--13888\relax
\mciteBstWouldAddEndPuncttrue
\mciteSetBstMidEndSepPunct{\mcitedefaultmidpunct}
{\mcitedefaultendpunct}{\mcitedefaultseppunct}\relax
\EndOfBibitem
\bibitem[Liu \emph{et~al.}(2014)Liu, Gall, and Khare]{Liu2014ElectronicPernitrides}
Z.~T.~Y. Liu, D.~Gall and S.~V. Khare, \emph{Physical Review B}, 2014, \textbf{90}, 134102\relax
\mciteBstWouldAddEndPuncttrue
\mciteSetBstMidEndSepPunct{\mcitedefaultmidpunct}
{\mcitedefaultendpunct}{\mcitedefaultseppunct}\relax
\EndOfBibitem
\bibitem[Bhadram \emph{et~al.}(2016)Bhadram, Kim, and Strobel]{Bhadram2016High-PressurePernitride}
V.~S. Bhadram, D.~Y. Kim and T.~A. Strobel, \emph{Chemistry of Materials}, 2016, \textbf{28}, 1616--1620\relax
\mciteBstWouldAddEndPuncttrue
\mciteSetBstMidEndSepPunct{\mcitedefaultmidpunct}
{\mcitedefaultendpunct}{\mcitedefaultseppunct}\relax
\EndOfBibitem
\bibitem[Chen \emph{et~al.}(2017)Chen, Wei, Chen, and Tian]{Chen2017TheoreticalPressure}
H.~Chen, J.~Wei, Y.~Chen and W.~Tian, \emph{Journal of Alloys and Compounds}, 2017, \textbf{726}, 1179--1185\relax
\mciteBstWouldAddEndPuncttrue
\mciteSetBstMidEndSepPunct{\mcitedefaultmidpunct}
{\mcitedefaultendpunct}{\mcitedefaultseppunct}\relax
\EndOfBibitem
\bibitem[Zhang \emph{et~al.}(2019)Zhang, Yan, and Wei]{Zhang2019UnexpectedHf}
M.~Zhang, H.~Yan and Q.~Wei, \emph{Journal of Alloys and Compounds}, 2019, \textbf{774}, 918--925\relax
\mciteBstWouldAddEndPuncttrue
\mciteSetBstMidEndSepPunct{\mcitedefaultmidpunct}
{\mcitedefaultendpunct}{\mcitedefaultseppunct}\relax
\EndOfBibitem
\bibitem[Niwa \emph{et~al.}(2019)Niwa, Yamamoto, Sasaki, and Hasegawa]{Niwa2019High-pressureDimer}
K.~Niwa, T.~Yamamoto, T.~Sasaki and M.~Hasegawa, \emph{Physical Review Materials}, 2019, \textbf{3}, 053601\relax
\mciteBstWouldAddEndPuncttrue
\mciteSetBstMidEndSepPunct{\mcitedefaultmidpunct}
{\mcitedefaultendpunct}{\mcitedefaultseppunct}\relax
\EndOfBibitem
\bibitem[Tsetseris \emph{et~al.}(2007)Tsetseris, Kalfagiannis, Logothetidis, and Pantelides]{Tsetseris2007StructureNitrides}
L.~Tsetseris, N.~Kalfagiannis, S.~Logothetidis and S.~T. Pantelides, \emph{Physical Review B}, 2007, \textbf{76}, 224107\relax
\mciteBstWouldAddEndPuncttrue
\mciteSetBstMidEndSepPunct{\mcitedefaultmidpunct}
{\mcitedefaultendpunct}{\mcitedefaultseppunct}\relax
\EndOfBibitem
\bibitem[Tsetseris \emph{et~al.}(2008)Tsetseris, Kalfagiannis, Logothetidis, and Pantelides]{Tsetseris2008TrappingStudy}
L.~Tsetseris, N.~Kalfagiannis, S.~Logothetidis and S.~T. Pantelides, \emph{Physical Review B}, 2008, \textbf{78}, 094111\relax
\mciteBstWouldAddEndPuncttrue
\mciteSetBstMidEndSepPunct{\mcitedefaultmidpunct}
{\mcitedefaultendpunct}{\mcitedefaultseppunct}\relax
\EndOfBibitem
\bibitem[Lee \emph{et~al.}(1996)Lee, Ho, Huang, Meletis, and Liu]{Lee1996HydrogenCoatings}
S.-C. Lee, W.-Y. Ho, C.-C. Huang, E.~Meletis and Y.~Liu, \emph{Journal of Materials Engineering and Performance}, 1996, \textbf{5}, 64--70\relax
\mciteBstWouldAddEndPuncttrue
\mciteSetBstMidEndSepPunct{\mcitedefaultmidpunct}
{\mcitedefaultendpunct}{\mcitedefaultseppunct}\relax
\EndOfBibitem
\bibitem[Ozsdolay \emph{et~al.}(2017)Ozsdolay, Balasubramanian, and Gall]{Ozsdolay2017CationNitride}
B.~Ozsdolay, K.~Balasubramanian and D.~Gall, \emph{Journal of Alloys and Compounds}, 2017, \textbf{705}, 631--637\relax
\mciteBstWouldAddEndPuncttrue
\mciteSetBstMidEndSepPunct{\mcitedefaultmidpunct}
{\mcitedefaultendpunct}{\mcitedefaultseppunct}\relax
\EndOfBibitem
\bibitem[Hu \emph{et~al.}(2015)Hu, Zhang, Gu, Huang, Zhang, Fan, Zhang, Wei, and Zheng]{Hu2015NegativeFilms}
C.~Hu, X.~Zhang, Z.~Gu, H.~Huang, S.~Zhang, X.~Fan, W.~Zhang, Q.~Wei and W.~Zheng, \emph{Scripta Materialia}, 2015, \textbf{108}, 141--146\relax
\mciteBstWouldAddEndPuncttrue
\mciteSetBstMidEndSepPunct{\mcitedefaultmidpunct}
{\mcitedefaultendpunct}{\mcitedefaultseppunct}\relax
\EndOfBibitem
\bibitem[Cetnar \emph{et~al.}(2018)Cetnar, Reed, Badescu, Vangala, Smith, and Look]{Cetnar2018ElectronicSubstrates}
J.~S. Cetnar, A.~N. Reed, S.~C. Badescu, S.~Vangala, H.~A. Smith and D.~C. Look, \emph{Applied Physics Letters}, 2018, \textbf{113}, \relax
\mciteBstWouldAddEndPuncttrue
\mciteSetBstMidEndSepPunct{\mcitedefaultmidpunct}
{\mcitedefaultendpunct}{\mcitedefaultseppunct}\relax
\EndOfBibitem
\bibitem[Kumagai \emph{et~al.}(2018)Kumagai, Tsunoda, and Oba]{Kumagai2018PointPrinciples}
Y.~Kumagai, N.~Tsunoda and F.~Oba, \emph{Physical Review Applied}, 2018, \textbf{9}, 034019\relax
\mciteBstWouldAddEndPuncttrue
\mciteSetBstMidEndSepPunct{\mcitedefaultmidpunct}
{\mcitedefaultendpunct}{\mcitedefaultseppunct}\relax
\EndOfBibitem
\bibitem[Deng \emph{et~al.}(2015)Deng, Ozsdolay, Zheng, Khare, and Gall]{Deng2015OpticalGap}
R.~Deng, B.~D. Ozsdolay, P.~Y. Zheng, S.~V. Khare and D.~Gall, \emph{Physical Review B}, 2015, \textbf{91}, 045104\relax
\mciteBstWouldAddEndPuncttrue
\mciteSetBstMidEndSepPunct{\mcitedefaultmidpunct}
{\mcitedefaultendpunct}{\mcitedefaultseppunct}\relax
\EndOfBibitem
\bibitem[Kerdsongpanya \emph{et~al.}(2012)Kerdsongpanya, Alling, and Eklund]{Kerdsongpanya2012EffectProperties}
S.~Kerdsongpanya, B.~Alling and P.~Eklund, \emph{Physical Review B}, 2012, \textbf{86}, 195140\relax
\mciteBstWouldAddEndPuncttrue
\mciteSetBstMidEndSepPunct{\mcitedefaultmidpunct}
{\mcitedefaultendpunct}{\mcitedefaultseppunct}\relax
\EndOfBibitem
\bibitem[Zhang \emph{et~al.}(2021)Zhang, Ghasemi, Koutn{\'{a}}, Xu, Gr{\"{u}}nst{\"{a}}udl, Song, Holec, He, Mayrhofer, and Bartosik]{Zhang2021CorrelatingFilms}
Z.~Zhang, A.~Ghasemi, N.~Koutn{\'{a}}, Z.~Xu, T.~Gr{\"{u}}nst{\"{a}}udl, K.~Song, D.~Holec, Y.~He, P.~H. Mayrhofer and M.~Bartosik, \emph{Materials {\&} Design}, 2021, \textbf{207}, 109844\relax
\mciteBstWouldAddEndPuncttrue
\mciteSetBstMidEndSepPunct{\mcitedefaultmidpunct}
{\mcitedefaultendpunct}{\mcitedefaultseppunct}\relax
\EndOfBibitem
\bibitem[Lee \emph{et~al.}(2012)Lee, Delley, Stampfl, and Soon]{Lee2012Environment-dependentVacancies}
T.~Lee, B.~Delley, C.~Stampfl and A.~Soon, \emph{Nanoscale}, 2012, \textbf{4}, 5183\relax
\mciteBstWouldAddEndPuncttrue
\mciteSetBstMidEndSepPunct{\mcitedefaultmidpunct}
{\mcitedefaultendpunct}{\mcitedefaultseppunct}\relax
\EndOfBibitem
\bibitem[Jhi \emph{et~al.}(2001)Jhi, Louie, Cohen, and Ihm]{Jhi2001VacancyNitrides}
S.-H. Jhi, S.~G. Louie, M.~L. Cohen and J.~Ihm, \emph{Physical Review Letters}, 2001, \textbf{86}, 3348--3351\relax
\mciteBstWouldAddEndPuncttrue
\mciteSetBstMidEndSepPunct{\mcitedefaultmidpunct}
{\mcitedefaultendpunct}{\mcitedefaultseppunct}\relax
\EndOfBibitem
\bibitem[Aissani \emph{et~al.}(2021)Aissani, Fellah, Chadli, Samad, Cheriet, Salhi, Nouveau, Wei{\ss}, Obrosov, and Alhussein]{Aissani2021InvestigatingSputtering}
L.~Aissani, M.~Fellah, A.~H. Chadli, M.~A. Samad, A.~Cheriet, F.~Salhi, C.~Nouveau, S.~Wei{\ss}, A.~Obrosov and A.~Alhussein, \emph{Journal of Materials Science}, 2021, \textbf{56}, 17319--17336\relax
\mciteBstWouldAddEndPuncttrue
\mciteSetBstMidEndSepPunct{\mcitedefaultmidpunct}
{\mcitedefaultendpunct}{\mcitedefaultseppunct}\relax
\EndOfBibitem
\bibitem[Holleck(1986)]{Holleck1986MaterialCoatings}
H.~Holleck, \emph{Journal of Vacuum Science {\&} Technology A: Vacuum, Surfaces, and Films}, 1986, \textbf{4}, 2661--2669\relax
\mciteBstWouldAddEndPuncttrue
\mciteSetBstMidEndSepPunct{\mcitedefaultmidpunct}
{\mcitedefaultendpunct}{\mcitedefaultseppunct}\relax
\EndOfBibitem
\bibitem[Jiang \emph{et~al.}(1991)Jiang, Wang, Schmidt, Dunlop, Haupt, and Gissler]{Jiang1991ElasticIndentation}
X.~Jiang, M.~Wang, K.~Schmidt, E.~Dunlop, J.~Haupt and W.~Gissler, \emph{Journal of Applied Physics}, 1991, \textbf{69}, 3053--3057\relax
\mciteBstWouldAddEndPuncttrue
\mciteSetBstMidEndSepPunct{\mcitedefaultmidpunct}
{\mcitedefaultendpunct}{\mcitedefaultseppunct}\relax
\EndOfBibitem
\bibitem[Karimi~Aghda \emph{et~al.}(2023)Karimi~Aghda, Bogdanovski, L{\"{o}}fler, Sua, Patterer, Holzapfel, le~Febvrier, Hans, Primetzhofer, and Schneider]{KarimiAghda2023ValenceNitrides}
S.~Karimi~Aghda, D.~Bogdanovski, L.~L{\"{o}}fler, H.~H. Sua, L.~Patterer, D.~M. Holzapfel, A.~le~Febvrier, M.~Hans, D.~Primetzhofer and J.~M. Schneider, \emph{Acta Materialia}, 2023, \textbf{255}, 119078\relax
\mciteBstWouldAddEndPuncttrue
\mciteSetBstMidEndSepPunct{\mcitedefaultmidpunct}
{\mcitedefaultendpunct}{\mcitedefaultseppunct}\relax
\EndOfBibitem
\bibitem[Zhang \emph{et~al.}(2016)Zhang, Balasubramanian, Ozsdolay, Mulligan, Khare, Zheng, and Gall]{Zhang2016GrowthMgO001}
K.~Zhang, K.~Balasubramanian, B.~Ozsdolay, C.~Mulligan, S.~Khare, W.~Zheng and D.~Gall, \emph{Surface and Coatings Technology}, 2016, \textbf{288}, 105--114\relax
\mciteBstWouldAddEndPuncttrue
\mciteSetBstMidEndSepPunct{\mcitedefaultmidpunct}
{\mcitedefaultendpunct}{\mcitedefaultseppunct}\relax
\EndOfBibitem
\bibitem[Ozsdolay \emph{et~al.}(2016)Ozsdolay, Mulligan, Balasubramanian, Huang, Khare, and Gall]{Ozsdolay2016CubicRatio}
B.~Ozsdolay, C.~Mulligan, K.~Balasubramanian, L.~Huang, S.~Khare and D.~Gall, \emph{Surface and Coatings Technology}, 2016, \textbf{304}, 98--107\relax
\mciteBstWouldAddEndPuncttrue
\mciteSetBstMidEndSepPunct{\mcitedefaultmidpunct}
{\mcitedefaultendpunct}{\mcitedefaultseppunct}\relax
\EndOfBibitem
\bibitem[Balasubramanian \emph{et~al.}(2017)Balasubramanian, Huang, and Gall]{Balasubramanian2017Phase1}
K.~Balasubramanian, L.~Huang and D.~Gall, \emph{Journal of Applied Physics}, 2017, \textbf{122}, \relax
\mciteBstWouldAddEndPuncttrue
\mciteSetBstMidEndSepPunct{\mcitedefaultmidpunct}
{\mcitedefaultendpunct}{\mcitedefaultseppunct}\relax
\EndOfBibitem
\bibitem[Balasubramanian \emph{et~al.}(2016)Balasubramanian, Khare, and Gall]{Balasubramanian2016Vacancy-inducedNitride}
K.~Balasubramanian, S.~Khare and D.~Gall, \emph{Physical Review B}, 2016, \textbf{94}, 174111\relax
\mciteBstWouldAddEndPuncttrue
\mciteSetBstMidEndSepPunct{\mcitedefaultmidpunct}
{\mcitedefaultendpunct}{\mcitedefaultseppunct}\relax
\EndOfBibitem
\bibitem[Browne \emph{et~al.}(1970)Browne, Liddell, Street, and Mills]{Browne1970AnCrN}
J.~D. Browne, P.~R. Liddell, R.~Street and T.~Mills, \emph{Phys. Status Solidi A}, 1970, \textbf{1}, 715--723\relax
\mciteBstWouldAddEndPuncttrue
\mciteSetBstMidEndSepPunct{\mcitedefaultmidpunct}
{\mcitedefaultendpunct}{\mcitedefaultseppunct}\relax
\EndOfBibitem
\bibitem[Ibberson and Cywinski(1992)]{Ibberson1992TheCrMoN}
R.~Ibberson and R.~Cywinski, \emph{Physica B: Condensed Matter}, 1992, \textbf{180-181}, 329--332\relax
\mciteBstWouldAddEndPuncttrue
\mciteSetBstMidEndSepPunct{\mcitedefaultmidpunct}
{\mcitedefaultendpunct}{\mcitedefaultseppunct}\relax
\EndOfBibitem
\bibitem[Subramanya~Herle \emph{et~al.}(1997)Subramanya~Herle, Hegde, Vasathacharya, Philip, Rama~Rao, and Sripathi]{SubramanyaHerle1997SynthesisCr2S3}
P.~Subramanya~Herle, M.~S. Hegde, N.~Y. Vasathacharya, S.~Philip, M.~V. Rama~Rao and T.~Sripathi, \emph{J. Solid State Chem.}, 1997, \textbf{134}, 120--127\relax
\mciteBstWouldAddEndPuncttrue
\mciteSetBstMidEndSepPunct{\mcitedefaultmidpunct}
{\mcitedefaultendpunct}{\mcitedefaultseppunct}\relax
\EndOfBibitem
\bibitem[Corliss \emph{et~al.}(1960)Corliss, Elliott, and Hastings]{Corliss1960AntiferromagneticCrN}
L.~M. Corliss, N.~Elliott and J.~M. Hastings, \emph{Physical Review}, 1960, \textbf{117}, 929--935\relax
\mciteBstWouldAddEndPuncttrue
\mciteSetBstMidEndSepPunct{\mcitedefaultmidpunct}
{\mcitedefaultendpunct}{\mcitedefaultseppunct}\relax
\EndOfBibitem
\bibitem[Engel \emph{et~al.}(1998)Engel, Schwarz, and Wolf]{Engel1998CorrosionDeposition}
P.~Engel, G.~Schwarz and G.~Wolf, \emph{Surface and Coatings Technology}, 1998, \textbf{98}, 1002--1007\relax
\mciteBstWouldAddEndPuncttrue
\mciteSetBstMidEndSepPunct{\mcitedefaultmidpunct}
{\mcitedefaultendpunct}{\mcitedefaultseppunct}\relax
\EndOfBibitem
\bibitem[Cunha \emph{et~al.}(1999)Cunha, Andritschky, Pischow, and Wang]{Cunha1999MicrostructureTechniques}
L.~Cunha, M.~Andritschky, K.~Pischow and Z.~Wang, \emph{Thin Solid Films}, 1999, \textbf{355-356}, 465--471\relax
\mciteBstWouldAddEndPuncttrue
\mciteSetBstMidEndSepPunct{\mcitedefaultmidpunct}
{\mcitedefaultendpunct}{\mcitedefaultseppunct}\relax
\EndOfBibitem
\bibitem[Tian \emph{et~al.}(2012)Tian, Xu, and Zhao]{Tian2012MicroscopicCrystals}
Y.~Tian, B.~Xu and Z.~Zhao, \emph{International Journal of Refractory Metals and Hard Materials}, 2012, \textbf{33}, 93--106\relax
\mciteBstWouldAddEndPuncttrue
\mciteSetBstMidEndSepPunct{\mcitedefaultmidpunct}
{\mcitedefaultendpunct}{\mcitedefaultseppunct}\relax
\EndOfBibitem
\bibitem[Kvashnin \emph{et~al.}(2017)Kvashnin, Oganov, Samtsevich, and Allahyari]{Kvashnin2017ComputationalMaterials}
A.~G. Kvashnin, A.~R. Oganov, A.~I. Samtsevich and Z.~Allahyari, \emph{The Journal of Physical Chemistry Letters}, 2017, \textbf{8}, 755--764\relax
\mciteBstWouldAddEndPuncttrue
\mciteSetBstMidEndSepPunct{\mcitedefaultmidpunct}
{\mcitedefaultendpunct}{\mcitedefaultseppunct}\relax
\EndOfBibitem
\bibitem[Mozafari \emph{et~al.}(2015)Mozafari, Alling, Steneteg, and Abrikosov]{Mozafari2015RolePrinciples}
E.~Mozafari, B.~Alling, P.~Steneteg and I.~A. Abrikosov, \emph{Physical Review B}, 2015, \textbf{91}, 094101\relax
\mciteBstWouldAddEndPuncttrue
\mciteSetBstMidEndSepPunct{\mcitedefaultmidpunct}
{\mcitedefaultendpunct}{\mcitedefaultseppunct}\relax
\EndOfBibitem
\bibitem[Tsuchiya \emph{et~al.}(1996)Tsuchiya, Kosuge, Ikeda, Shigematsu, Yamaguchi, and Nakayama]{Tsuchiya1996Non-stoichiometrySputtering}
Y.~Tsuchiya, K.~Kosuge, Y.~Ikeda, T.~Shigematsu, S.~Yamaguchi and N.~Nakayama, \emph{Materials Transactions, JIM}, 1996, \textbf{37}, 121--129\relax
\mciteBstWouldAddEndPuncttrue
\mciteSetBstMidEndSepPunct{\mcitedefaultmidpunct}
{\mcitedefaultendpunct}{\mcitedefaultseppunct}\relax
\EndOfBibitem
\bibitem[Inumaru \emph{et~al.}(2007)Inumaru, Koyama, Imo-oka, and Yamanaka]{Inumaru2007ControllingGrowth}
K.~Inumaru, K.~Koyama, N.~Imo-oka and S.~Yamanaka, \emph{Physical Review B}, 2007, \textbf{75}, 054416\relax
\mciteBstWouldAddEndPuncttrue
\mciteSetBstMidEndSepPunct{\mcitedefaultmidpunct}
{\mcitedefaultendpunct}{\mcitedefaultseppunct}\relax
\EndOfBibitem
\bibitem[Rojas and Ulloa(2018)]{Rojas2018EnergeticsCrN}
T.~Rojas and S.~E. Ulloa, \emph{Physical Review B}, 2018, \textbf{98}, 214111\relax
\mciteBstWouldAddEndPuncttrue
\mciteSetBstMidEndSepPunct{\mcitedefaultmidpunct}
{\mcitedefaultendpunct}{\mcitedefaultseppunct}\relax
\EndOfBibitem
\bibitem[Suzuki \emph{et~al.}(2006)Suzuki, Inoue, Saito, Hirai, Suematsu, Jiang, and Yatsui]{Suzuki2006InfluenceDeposition}
T.~Suzuki, J.~Inoue, H.~Saito, M.~Hirai, H.~Suematsu, W.~Jiang and K.~Yatsui, \emph{Thin Solid Films}, 2006, \textbf{515}, 2161--2166\relax
\mciteBstWouldAddEndPuncttrue
\mciteSetBstMidEndSepPunct{\mcitedefaultmidpunct}
{\mcitedefaultendpunct}{\mcitedefaultseppunct}\relax
\EndOfBibitem
\bibitem[Du \emph{et~al.}(2021)Du, Chen, Chen, and Hu]{Du2021InfluenceCoating}
J.~W. Du, L.~Chen, J.~Chen and C.~Hu, \emph{Surface and Coatings Technology}, 2021, \textbf{411}, 126992\relax
\mciteBstWouldAddEndPuncttrue
\mciteSetBstMidEndSepPunct{\mcitedefaultmidpunct}
{\mcitedefaultendpunct}{\mcitedefaultseppunct}\relax
\EndOfBibitem
\bibitem[Zhao \emph{et~al.}(2016)Zhao, Bao, Tian, Duan, Liu, and Cui]{Zhao2016PotentiallyPressure}
Z.~Zhao, K.~Bao, F.~Tian, D.~Duan, B.~Liu and T.~Cui, \emph{Physical Review B}, 2016, \textbf{93}, 214104\relax
\mciteBstWouldAddEndPuncttrue
\mciteSetBstMidEndSepPunct{\mcitedefaultmidpunct}
{\mcitedefaultendpunct}{\mcitedefaultseppunct}\relax
\EndOfBibitem
\bibitem[Liu \emph{et~al.}(2019)Liu, Li, Liu, Wei, Tian, Bao, Duan, Liu, and Cui]{Liu2019TheStudy}
Y.~Liu, D.~Li, Z.~Liu, S.~Wei, F.~Tian, K.~Bao, D.~Duan, B.~Liu and T.~Cui, \emph{Computational Materials Science}, 2019, \textbf{158}, 282--288\relax
\mciteBstWouldAddEndPuncttrue
\mciteSetBstMidEndSepPunct{\mcitedefaultmidpunct}
{\mcitedefaultendpunct}{\mcitedefaultseppunct}\relax
\EndOfBibitem
\bibitem[Kanyanta(2016)]{Kanyanta2016HardOverview}
V.~Kanyanta, \emph{Microstructure-Property Correlations for Hard, Superhard, and Ultrahard Materials}, Springer International Publishing, Cham, 2016, pp. 1--23\relax
\mciteBstWouldAddEndPuncttrue
\mciteSetBstMidEndSepPunct{\mcitedefaultmidpunct}
{\mcitedefaultendpunct}{\mcitedefaultseppunct}\relax
\EndOfBibitem
\bibitem[Kresse and Hafner(1993)]{Kresse1993iAbMetals}
G.~Kresse and J.~Hafner, \emph{Physical Review B}, 1993, \textbf{47}, 558--561\relax
\mciteBstWouldAddEndPuncttrue
\mciteSetBstMidEndSepPunct{\mcitedefaultmidpunct}
{\mcitedefaultendpunct}{\mcitedefaultseppunct}\relax
\EndOfBibitem
\bibitem[Kresse and Furthm{\"{u}}ller(1996)]{Kresse1996EfficiencySet}
G.~Kresse and J.~Furthm{\"{u}}ller, \emph{Computational Materials Science}, 1996, \textbf{6}, 15--50\relax
\mciteBstWouldAddEndPuncttrue
\mciteSetBstMidEndSepPunct{\mcitedefaultmidpunct}
{\mcitedefaultendpunct}{\mcitedefaultseppunct}\relax
\EndOfBibitem
\bibitem[Perdew \emph{et~al.}(1996)Perdew, Burke, and Ernzerhof]{Perdew1996GeneralizedSimple}
J.~P. Perdew, K.~Burke and M.~Ernzerhof, \emph{Physical Review Letters}, 1996, \textbf{77}, 3865--3868\relax
\mciteBstWouldAddEndPuncttrue
\mciteSetBstMidEndSepPunct{\mcitedefaultmidpunct}
{\mcitedefaultendpunct}{\mcitedefaultseppunct}\relax
\EndOfBibitem
\bibitem[Langreth and Mehl(1983)]{Langreth1983BeyondProperties}
D.~C. Langreth and M.~J. Mehl, \emph{Physical Review B}, 1983, \textbf{28}, 1809--1834\relax
\mciteBstWouldAddEndPuncttrue
\mciteSetBstMidEndSepPunct{\mcitedefaultmidpunct}
{\mcitedefaultendpunct}{\mcitedefaultseppunct}\relax
\EndOfBibitem
\bibitem[Bl{\"{o}}chl(1994)]{Blochl1994ProjectorMethod}
P.~E. Bl{\"{o}}chl, \emph{Physical Review B}, 1994, \textbf{50}, 17953--17979\relax
\mciteBstWouldAddEndPuncttrue
\mciteSetBstMidEndSepPunct{\mcitedefaultmidpunct}
{\mcitedefaultendpunct}{\mcitedefaultseppunct}\relax
\EndOfBibitem
\bibitem[Anisimov \emph{et~al.}(1991)Anisimov, Zaanen, and Andersen]{Anisimov1991BandiI/i}
V.~I. Anisimov, J.~Zaanen and O.~K. Andersen, \emph{Physical Review B}, 1991, \textbf{44}, 943--954\relax
\mciteBstWouldAddEndPuncttrue
\mciteSetBstMidEndSepPunct{\mcitedefaultmidpunct}
{\mcitedefaultendpunct}{\mcitedefaultseppunct}\relax
\EndOfBibitem
\bibitem[Methfessel and Paxton(1989)]{Methfessel1989High-precisionMetals}
M.~Methfessel and A.~T. Paxton, \emph{Physical Review B}, 1989, \textbf{40}, 3616--3621\relax
\mciteBstWouldAddEndPuncttrue
\mciteSetBstMidEndSepPunct{\mcitedefaultmidpunct}
{\mcitedefaultendpunct}{\mcitedefaultseppunct}\relax
\EndOfBibitem
\bibitem[Wang \emph{et~al.}(2021)Wang, Xu, Liu, Tang, and Geng]{Wang2021VASPKIT:Code}
V.~Wang, N.~Xu, J.-C. Liu, G.~Tang and W.-T. Geng, \emph{Computer Physics Communications}, 2021, \textbf{267}, 108033\relax
\mciteBstWouldAddEndPuncttrue
\mciteSetBstMidEndSepPunct{\mcitedefaultmidpunct}
{\mcitedefaultendpunct}{\mcitedefaultseppunct}\relax
\EndOfBibitem
\bibitem[Born and Huang(1996)]{Born1996DynamicalLattices}
M.~Born and K.~Huang, \emph{{Dynamical Theory Of Crystal Lattices}}, Oxford University PressNew York, NY, 1996\relax
\mciteBstWouldAddEndPuncttrue
\mciteSetBstMidEndSepPunct{\mcitedefaultmidpunct}
{\mcitedefaultendpunct}{\mcitedefaultseppunct}\relax
\EndOfBibitem
\bibitem[Mouhat and Coudert(2014)]{Mouhat2014NecessarySystems}
F.~Mouhat and F.-X. Coudert, \emph{Physical Review B}, 2014, \textbf{90}, 224104\relax
\mciteBstWouldAddEndPuncttrue
\mciteSetBstMidEndSepPunct{\mcitedefaultmidpunct}
{\mcitedefaultendpunct}{\mcitedefaultseppunct}\relax
\EndOfBibitem
\bibitem[Grau-Crespo \emph{et~al.}(2007)Grau-Crespo, Hamad, Catlow, and Leeuw]{Grau-Crespo2007Symmetry-adaptedSolids}
R.~Grau-Crespo, S.~Hamad, C.~R.~A. Catlow and N.~H.~d. Leeuw, \emph{Journal of Physics: Condensed Matter}, 2007, \textbf{19}, 256201\relax
\mciteBstWouldAddEndPuncttrue
\mciteSetBstMidEndSepPunct{\mcitedefaultmidpunct}
{\mcitedefaultendpunct}{\mcitedefaultseppunct}\relax
\EndOfBibitem
\bibitem[Hirel(2015)]{Hirel2015Atomsk:Files}
P.~Hirel, \emph{Computer Physics Communications}, 2015, \textbf{197}, 212--219\relax
\mciteBstWouldAddEndPuncttrue
\mciteSetBstMidEndSepPunct{\mcitedefaultmidpunct}
{\mcitedefaultendpunct}{\mcitedefaultseppunct}\relax
\EndOfBibitem
\bibitem[Aroyo \emph{et~al.}(2006)Aroyo, Kirov, Capillas, Perez-Mato, and Wondratschek]{Aroyo2006BilbaoGroups}
M.~I. Aroyo, A.~Kirov, C.~Capillas, J.~M. Perez-Mato and H.~Wondratschek, \emph{Acta Crystallographica Section A Foundations of Crystallography}, 2006, \textbf{62}, 115--128\relax
\mciteBstWouldAddEndPuncttrue
\mciteSetBstMidEndSepPunct{\mcitedefaultmidpunct}
{\mcitedefaultendpunct}{\mcitedefaultseppunct}\relax
\EndOfBibitem
\bibitem[Zhang and Northrup(1991)]{Zhang1991ChemicalSelf-diffusion}
S.~Zhang and J.~Northrup, \emph{Physical Review Letters}, 1991, \textbf{67}, 2339--2342\relax
\mciteBstWouldAddEndPuncttrue
\mciteSetBstMidEndSepPunct{\mcitedefaultmidpunct}
{\mcitedefaultendpunct}{\mcitedefaultseppunct}\relax
\EndOfBibitem
\bibitem[Kr{\"{o}}ger and Vink(1956)]{Kroger1956RelationsSolids}
F.~Kr{\"{o}}ger and H.~Vink, \emph{H.~Vink}, 1956, pp. 307--435\relax
\mciteBstWouldAddEndPuncttrue
\mciteSetBstMidEndSepPunct{\mcitedefaultmidpunct}
{\mcitedefaultendpunct}{\mcitedefaultseppunct}\relax
\EndOfBibitem
\bibitem[Buckeridge \emph{et~al.}(2014)Buckeridge, Scanlon, Walsh, and Catlow]{Buckeridge2014AutomatedCompounds}
J.~Buckeridge, D.~Scanlon, A.~Walsh and C.~Catlow, \emph{Computer Physics Communications}, 2014, \textbf{185}, 330--338\relax
\mciteBstWouldAddEndPuncttrue
\mciteSetBstMidEndSepPunct{\mcitedefaultmidpunct}
{\mcitedefaultendpunct}{\mcitedefaultseppunct}\relax
\EndOfBibitem
\bibitem[Kumagai and Oba(2014)]{Kumagai2014Electrostatics-basedCalculations}
Y.~Kumagai and F.~Oba, \emph{Physical Review B}, 2014, \textbf{89}, 195205\relax
\mciteBstWouldAddEndPuncttrue
\mciteSetBstMidEndSepPunct{\mcitedefaultmidpunct}
{\mcitedefaultendpunct}{\mcitedefaultseppunct}\relax
\EndOfBibitem
\bibitem[Arrigoni and Madsen(2021)]{Arrigoni2021Spinney:Python}
M.~Arrigoni and G.~K. Madsen, \emph{Computer Physics Communications}, 2021, \textbf{264}, 107946\relax
\mciteBstWouldAddEndPuncttrue
\mciteSetBstMidEndSepPunct{\mcitedefaultmidpunct}
{\mcitedefaultendpunct}{\mcitedefaultseppunct}\relax
\EndOfBibitem
\bibitem[Freysoldt \emph{et~al.}(2009)Freysoldt, Neugebauer, and Van~de Walle]{Freysoldt2009FullyCalculations}
C.~Freysoldt, J.~Neugebauer and C.~G. Van~de Walle, \emph{Physical Review Letters}, 2009, \textbf{102}, 016402\relax
\mciteBstWouldAddEndPuncttrue
\mciteSetBstMidEndSepPunct{\mcitedefaultmidpunct}
{\mcitedefaultendpunct}{\mcitedefaultseppunct}\relax
\EndOfBibitem
\bibitem[Freysoldt \emph{et~al.}(2011)Freysoldt, Neugebauer, and Van~de Walle]{Freysoldt2011ElectrostaticSupercells}
C.~Freysoldt, J.~Neugebauer and C.~G. Van~de Walle, \emph{physica status solidi (b)}, 2011, \textbf{248}, 1067--1076\relax
\mciteBstWouldAddEndPuncttrue
\mciteSetBstMidEndSepPunct{\mcitedefaultmidpunct}
{\mcitedefaultendpunct}{\mcitedefaultseppunct}\relax
\EndOfBibitem
\bibitem[Freysoldt \emph{et~al.}(2014)Freysoldt, Grabowski, Hickel, Neugebauer, Kresse, Janotti, and Van~de Walle]{Freysoldt2014First-principlesSolids}
C.~Freysoldt, B.~Grabowski, T.~Hickel, J.~Neugebauer, G.~Kresse, A.~Janotti and C.~G. Van~de Walle, \emph{Reviews of Modern Physics}, 2014, \textbf{86}, 253--305\relax
\mciteBstWouldAddEndPuncttrue
\mciteSetBstMidEndSepPunct{\mcitedefaultmidpunct}
{\mcitedefaultendpunct}{\mcitedefaultseppunct}\relax
\EndOfBibitem
\bibitem[Yan and Chen(2014)]{Yan2014StructuralStudy}
M.~Yan and H.~Chen, \emph{Computational Materials Science}, 2014, \textbf{88}, 81--85\relax
\mciteBstWouldAddEndPuncttrue
\mciteSetBstMidEndSepPunct{\mcitedefaultmidpunct}
{\mcitedefaultendpunct}{\mcitedefaultseppunct}\relax
\EndOfBibitem
\bibitem[Wang \emph{et~al.}(2022)Wang, Tu, Yuan, Qian, Jonnard, and Lan]{Wang2022CompositionFilms}
J.~Wang, Y.~Tu, Y.~Yuan, H.~Qian, P.~Jonnard and R.~Lan, \emph{Surface and Interface Analysis}, 2022, \textbf{54}, 1142--1150\relax
\mciteBstWouldAddEndPuncttrue
\mciteSetBstMidEndSepPunct{\mcitedefaultmidpunct}
{\mcitedefaultendpunct}{\mcitedefaultseppunct}\relax
\EndOfBibitem
\bibitem[Pankove \emph{et~al.}(1991)Pankove, Johnson, and Nickel]{Pankove1991HydrogenSemiconductors}
J.~I. Pankove, N.~M. Johnson and N.~H. Nickel, author, 1991\relax
\mciteBstWouldAddEndPuncttrue
\mciteSetBstMidEndSepPunct{\mcitedefaultmidpunct}
{\mcitedefaultendpunct}{\mcitedefaultseppunct}\relax
\EndOfBibitem
\bibitem[Pearton \emph{et~al.}(1992)Pearton, Corbett, and Stavola]{Pearton1992HydrogenSemiconductors}
S.~J. Pearton, J.~W. Corbett and M.~Stavola, \emph{{Hydrogen in Crystalline Semiconductors}}, Springer Berlin Heidelberg, Berlin, Heidelberg, 1992, vol.~16\relax
\mciteBstWouldAddEndPuncttrue
\mciteSetBstMidEndSepPunct{\mcitedefaultmidpunct}
{\mcitedefaultendpunct}{\mcitedefaultseppunct}\relax
\EndOfBibitem
\bibitem[Pettinari \emph{et~al.}(2014)Pettinari, Felici, Trotta, Capizzi, and Polimeni]{Pettinari2014HydrogenNanostructuring}
G.~Pettinari, M.~Felici, R.~Trotta, M.~Capizzi and A.~Polimeni, \emph{Journal of Applied Physics}, 2014, \textbf{115}, \relax
\mciteBstWouldAddEndPuncttrue
\mciteSetBstMidEndSepPunct{\mcitedefaultmidpunct}
{\mcitedefaultendpunct}{\mcitedefaultseppunct}\relax
\EndOfBibitem
\bibitem[Zhou \emph{et~al.}(2014)Zhou, K{\"{o}}rmann, Holec, Bartosik, Grabowski, Neugebauer, and Mayrhofer]{Zhou2014StructuralExperiment}
L.~Zhou, F.~K{\"{o}}rmann, D.~Holec, M.~Bartosik, B.~Grabowski, J.~Neugebauer and P.~H. Mayrhofer, \emph{Physical Review B}, 2014, \textbf{90}, 184102\relax
\mciteBstWouldAddEndPuncttrue
\mciteSetBstMidEndSepPunct{\mcitedefaultmidpunct}
{\mcitedefaultendpunct}{\mcitedefaultseppunct}\relax
\EndOfBibitem
\bibitem[Zhang \emph{et~al.}(2020)Zhang, Chen, Holec, Liebscher, Koutn{\'{a}}, Bartosik, Zheng, Dehm, and Mayrhofer]{Zhang2020MappingScale}
Z.~Zhang, Z.~Chen, D.~Holec, C.~H. Liebscher, N.~Koutn{\'{a}}, M.~Bartosik, Y.~Zheng, G.~Dehm and P.~H. Mayrhofer, \emph{Acta Materialia}, 2020, \textbf{194}, 343--353\relax
\mciteBstWouldAddEndPuncttrue
\mciteSetBstMidEndSepPunct{\mcitedefaultmidpunct}
{\mcitedefaultendpunct}{\mcitedefaultseppunct}\relax
\EndOfBibitem
\bibitem[Wang \emph{et~al.}(2016)Wang, Yu, Zhang, Wang, Leinenweber, He, and Zhao]{Wang2016SynthesisCrN}
S.~Wang, X.~Yu, J.~Zhang, L.~Wang, K.~Leinenweber, D.~He and Y.~Zhao, \emph{Crystal Growth {\&} Design}, 2016, \textbf{16}, 351--358\relax
\mciteBstWouldAddEndPuncttrue
\mciteSetBstMidEndSepPunct{\mcitedefaultmidpunct}
{\mcitedefaultendpunct}{\mcitedefaultseppunct}\relax
\EndOfBibitem
\end{mcitethebibliography}

\end{document}